\documentclass[journal]{IEEEtran}
\usepackage{graphicx}
\usepackage{amsmath}
\usepackage{algorithm}
\usepackage{hyperref}
\usepackage{booktabs} 
\usepackage{tabularx} 

\usepackage{booktabs} 
\usepackage{tabularx} 
\usepackage{caption}  
\usepackage{mathtools}
\usepackage{amssymb}
\usepackage{amsmath}
\usepackage{multirow}

\usepackage{array}
\usepackage{makecell}
\usepackage{textgreek}
\usepackage{comment}
\usepackage{bm}
\usepackage{booktabs}
\usepackage{color}
\usepackage{tabularx}
\usepackage{algorithmicx}
\usepackage{algpseudocode}
\usepackage{subcaption}
\usepackage[normalem]{ulem}
\newtheorem{theorem}{Theorem}

\algnewcommand\Input{\item[\textbf{Input:}]}  
\algnewcommand\Output{\item[\textbf{Output:}]}  
\usepackage{graphicx} 
\begin{document}


\title{First Glimpse on Physical Layer Security in Internet of Vehicles: Transformed from Communication Interference to Sensing Interference}
 
\author{Kaixuan Li, Kan Yu,~\IEEEmembership{Member,~IEEE}, Xiaowu Liu, Qixun Zhang,~\IEEEmembership{Member,~IEEE}, Zhiyong Feng,~\IEEEmembership{Senior Member,~IEEE}, and Dong Li,~\IEEEmembership{Senior Member,~IEEE}
\thanks{This work is supported by the National Natural Science Foundation of China with Grant 62301076, the Macao Young Scholars Program with Grant AM2023015, Fundamental Research Funds for the Central Universities with Grant  24820232023YQTD01, National Natural Science Foundation of China with Grants 62341101 and 62321001, Beijing Municipal Natural Science Foundation with Grant L232003, and National Key Research and Development Program of China with Grant 2022YFB4300403, and the Science and Technology Development Fund, Macau SAR, under Grant 0188/2023/RIA3.
}

\thanks{K. Li is with the School of Computer Science, Qufu Normal University, Rizhao, P.R. China. E-mail: lkx0311@126.com;}
\thanks{K. Yu (\emph{the corresponding author}) is with the School of Computer Science and Engineering, Macau University of Science and Technology, Taipa, Macau, 999078, P. R. China;
the Key Laboratory of Universal Wireless Communications, Ministry of Education, Beijing University of Posts and Telecommunications, Beijing, 100876, P.R. China. E-mail: kanyu1108@126.com;}
\thanks{X. Liu is with the School of Computer Science, Qufu Normal University, Rizhao, P.R. China. E-mail: liuxw@qfnu.edu.cn;}
\thanks{Q. Zhang is with the Key Laboratory of Universal
Wireless Communications, Ministry of Education, Beijing University of Posts and Telecommunications, Beijing, 100876, P.R. China. E-mail: zhangqixun@bupt.edu.cn;}
\thanks{Z. Feng is with the Key Laboratory of Universal Wireless Communications, Ministry of Education, Beijing University of Posts and Telecommunications, Beijing, 100876, P.R. China. E-mail: fengzy@bupt.edu.cn;}
\thanks{D. Li is with the School of Computer Science and Engineering, Macau University of Science and Technology, Taipa, Macau, China. E-mail: dli@must.edu.mo.}


}

\markboth{IEEE Transactions on Communications,~Vol.~, No.~, 2024}%
{Shell \Baogui Huang{\textit{et al.}}: Shortest Link Scheduling Under SINR}
\maketitle
\begin{abstract}
Integrated sensing and communication (ISAC) plays a crucial role in the Internet of Vehicles (IoV), serving as a key factor in enhancing driving safety and traffic efficiency. To address the security challenges of the confidential information transmission caused by the inherent openness nature of wireless medium, different from current physical layer security (PLS) methods, which depends on the \emph{additional communication interference} costing extra power resources, in this paper, we investigate a novel PLS solution, under which the \emph{inherent radar sensing interference} of the vehicles is utilized to secure wireless communications. To measure the performance of PLS methods in ISAC-based IoV systems, we first define an improved security performance metric called by transmission reliability and sensing accuracy based secrecy rate (TRSA\_SR), and derive closed-form expressions of connection outage probability (COP), secrecy outage probability (SOP), success ranging probability (SRP) for evaluating transmission reliability, security and sensing accuracy, respectively. Furthermore, we formulate an optimization problem to maximize the TRSA\_SR by utilizing radar sensing interference and joint design of the communication duration, transmission power and straight trajectory of the legitimate transmitter. Finally, the non-convex feature of formulated problem is solved through the problem decomposition and alternating optimization.  
Simulations indicate that compared with traditional PLS methods obtaining a non-positive STC, the proposed method achieves a secrecy rate of 3.92bps/Hz for different settings of noise power.

\end{abstract}

\IEEEpeerreviewmaketitle

\section{Introduction}\label{sec:introduction}

The IoV, equipped with ISAC capabilities, serves as a significant enabler of intelligent transportation systems (ITS). Driven by the demands for high-precision sensing, ultra-low communication latency, and ultra-high data rate, 
the transition of communication and sensing technologies towards millimeter-wave (mmWave) frequency bands is an irresistible general trend \cite{Que2023Joint,Qi2024Multiuse}. Due to the inherent openness nature of the wireless medium, strong directional beamforming (BF) of mmWave, and vehicles' mobility, the confidentiality and security of information transmission in IoV systems become extremely challenging. For example, if the BF of the confidential information is aimed at the eavesdroppers (Eves), there will be no security at all. Consequently, the IoV faces pressing security challenges that require immediate attention. 

Aiming at maximizing difference effects between legitimate channel and eavesdropping channel, PLS emerges as an effective supplement of traditional encryption methods to address security challenges by utilizing the inherent unpredictability of wireless channels with a lower complexity satisfying the features of IoV \cite{shannon1949communication,wyner1975wire}. In fact, with the help of communication interference, the core idea behind PLS implementation is to suppress the quality of the eavesdropping channel to be inferior to that of the legitimate channel.
However, these PLS methods cannot be directly extended to secure wireless communications in
ISAC-assisted IoV systems,
since communication-sensing coupled interference exists rather than only communication interference when the communication and sensing functions work in the same or similar frequency band, along with the strong directional BF and vehicles' mobility in the context of ISAC-based IoV scenarios. Against this background, new effective PLS methods urgently need to be proposed for satisfying requirements of the communication, sensing, and security in ISAC-based IoV systems.

The sensing interference is generally considered harmful to communication systems, and transceiver design \cite{Shlezinger2021DeepSIC}, BF design \cite{Liu2019Multi} and collaboration \cite{Li2022JointT} are popular methods to deal with the sensing interference for mitigating the effects on transmission reliability and sensing accuracy. From the perspective of PLS design, similar to communication interference, it can be also used to protect the confidential information from being eavesdropped. In addition, strong directional sensing BF potentially provides an effective means to only suppress the quality of wiretap channel rather than legitimate channel.
Then a natural question arises: \emph{\textbf{``is it possible and how to make a transformation from `communication interference based security' to `sensing interference based security'?''}}. This problem remains unclear and is not solved in existing works regarding the traditional PLS.

To demonstrate the rationality and validity of above transformation, in the context of an ISAC based IoV scenario, where a transmitting vehicle (Alice), a receiving vehicle (Bob), multiple moving vehicles (Carols) monitoring the target by using sensing capabilities, and an Eve are considered, in this paper, we study the joint design of the transmission power and straight trajectory of the Alice, and utilize the strong sensing BF of Carols to suppress the quality of eavesdropping channel, while satisfying the expected sensing accuracy of Carols, transmission reliability and security of Bob.
To the best of our knowledge, this is the first paper to only utilize sensing interference for secure wireless communications. The main contributions of this paper can be summarized as follows.
\begin{itemize}
    \item To address communication-sensing coupled interference, we derive closed-form expressions for connection outage probability (COP), secrecy outage probability (SOP), and success range probability (SRP), which offer significant insights into how key system parameters influence sensing accuracy, transmission reliability, and security; 
    \item To achieve the desired communication and sensing performance under constraints from COP, SOP, and SRP, we introduce a transmission reliability and sensing accuracy-based secrecy rate (TRSA\_SR) metric. We formulate an optimization problem to maximize TRSA\_SR through joint design of transmission time (for sensing interference utilization), power allocation, and the trajectory of Alice;
    \item Given the non-convex nature of the optimization problem, an alternating algorithm is proposed to iteratively optimize Alice’s transmission power allocation and trajectory. Additionally, we provide detailed analyses of the algorithm’s time complexity and convergence;
    \item Simulation results validate the effectiveness of sensing interference based PLS method. Notably, when traditional PLS approaches fail to achieve a positive TRSA\_SR, our proposed scheme achieves a TRSA\_SR of 3.92 bps/Hz for different settings of noise power.
\end{itemize}

The remainder of the paper is organized as follows: The related works are demonstrated in Section \ref{sec:related work}. In Section \ref{sec:network model}, the network model and performance metrics are presented, then the optimization for secrecy rate maximization is formulated. In Section \ref{sec:opt}, closed-form expressions of COP, SOP and SRP are derived at first, then communication duration of confidential information is optimized, and transmit power optimization and trajectory optimization are discussed, respectively. Simulations demonstrate the effectiveness of the proposed scheme in Section \ref{sec:simulation}. Finally, conclusions and future works are discussed in Section \ref{sec:conclusion}.

\begin{table*}[t]  
    \caption{A Brief summary of the current PLS methods considering reliability, sensing and security}
    \centering
    \label{tab:methods}
    \begin{tabular}{|>{\centering\arraybackslash}m{1.5cm}|>{\centering\arraybackslash}m{3.5cm}|>{\centering\arraybackslash}m{2cm}|>{\centering\arraybackslash}m{1.0cm}|>{\centering\arraybackslash}m{1.0cm}|>{\centering\arraybackslash}m{5.2cm}|}  
        \hline  
        \textbf{Reference}   & \textbf{Core ideas of PLS design}     & \textbf{Communication index}   & \textbf{Sensing index}  & \textbf{Security index}  & \textbf{Summaries} \\ \hline 
        \cite{wang2020physical}    & Joint design AN power and legitimate signal's BF    & $\checkmark$    & $\times$ & $\checkmark$ & \multirow{4}{5.2cm}{\centering 1) Secrecy enhancement empowered by communication interference; \\2) Extra energy consumption due to the generation of additional AN signals; \\3) Without considering the effects of communication-sensing coupled interference on the performance of communication, sensing, and security} \\ \cline{1-5}
        \cite{Zhang2019Transmit,yin2021uav} & Joint design of AN's power, legitimate signal's BF & $\checkmark$ & $\times$ & $\times$  & \\ \cline{1-5}
        \cite{Li2019UAV,Xu2021Low} & Joint design of AN's power and UAV's trajectory  & $\checkmark$ & $\times$ & $\times$  & \\ \cline{1-5}
        \cite{Zhou2018Improving} & Joint design of AN's power and UAV's trajectory  & $\checkmark$ & $\times$ & $\checkmark$  & \\ \cline{1-5}
         \cite{Chen2024Physical} & RIS/BF-based scenario & $\checkmark$ & $\times$ & $\checkmark$  & \\ \hline
         
        \cite{Liu2022Dynamic,Li2022Optimization} & $\times$ & $\times$ & $\times$ & $\times$  &  Optimized algorithms and driving safety \\ \hline
         \cite{Fang2020Stochastic,Zhang2024Coexistence} & $\times$& $\times$ & $\checkmark$   & $\times$ & without considering PLS \\ \hline

          \cite{chu2023joint} & Sensing signal-assisted BF, or AN-assisted BF & $\checkmark$ & $\checkmark$   & $\times$ & 1) PLS design based on sensing/AN signals; 2) Without considering the mobility of vehicles \\ \hline
          
         Method in this paper & \textbf{\emph{Sensing interference empowered joint design of transmission power and trajectory of vehicles}} & $\checkmark$ & $\checkmark$  & $\checkmark$ &
        1) PLS design based on sensing interference;
        2) Considering the vehicles' mobility, trajectory and moving time\\[1.2em] \hline
    \end{tabular}
\end{table*}
\section{Related Works}\label{sec:related work}


Wireless PLS technologies become particularly critical in the IoV, in terms of secure autonomous driving, vehicle-to-infrastructure. Concurrently, many efforts on the PLS design focused on the communication interference utilization by considering different legitimate and eavesdropping channels, such as the artificial noise (AN) scheme, joint design of the phase shift matrix and transmission power of reconfigurable intelligent surface (RIS), and the joint design of the transmission power and trajectory of the UAV.

\subsection{The communication interference based PLS in IoV systems}
AN scheme is one of most classic PLS techniques to effectively reduce the quality of eavesdropping channel, thereby enhancing the security.
In particular, \emph{Wang et al.} explored PLS in cellular vehicular networks, under which the legitimate user transmits confidential information while generating AN signals simultaneously for enhancing communication security \cite{wang2020physical}. Similarly, by design optimal AN-aided BF, \emph{Zhang et al.} proposed a Layered PLS model that minimizes transmission power while satisfying secrecy rate requirements \cite{Zhang2019Transmit}. The importance of attracting AN to the PLS were also emphasized in \cite{yin2021uav,jin2024enhanced}.
Featured by establishing line-of-sight (LoS) channels, the UAV and RIS provide new potential advantages in suppressing the wiretap channel. On the basis of this idea, in \cite{Li2019UAV}, \emph{Li et al.} regarded moving UAVs as jammers to generate AN signals for interfering with Eves, and jointly optimized the UAVs' trajectory and transmission power to enhance the security. \emph{Chen et al.} utilized RIS-assisted V2V communications and derived the upper bound of the secrecy capacity and the approximate expression of the secrecy outage probability under Rayleigh fading channel \cite{Chen2024Physical}.
It was demonstrated that optimizing UAV trajectory can effectively improve communication performance in \cite{Xu2021Low,Zhou2018Improving}.

However, in the context of ISAC-enabled IoV systems, communication and sensing frequency bands are often close, overlapping, or even identical, which results in communication-sensing coupled interference. 
Therefore, above PLS methods centered on communication interference cannot be directly extended to the secure information transmission in IoV systems featured by coupled interference. Moreover, although various vehicles trajectory optimization schemes, such as \cite{Liu2022Dynamic, Li2022Optimization}, have been proposed for V2V communication, studies specifically leveraging vehicle trajectory optimization to secure wireless communications remain relatively scarce.

In \cite{chu2023joint}, \emph{ Chu et al.} considered the design of 
PLS methods in an ISAC-based system. With the assumption of Eves' perfect CSI known beforehand, strong radar sensing signals were used to suppress the eavesdropping channel, then joint optimization of secure transmission BF and radar receiving filters was done for enhancing the system security. Otherwise, 
all available power resources of the base station and radar were utilized to design BF matrix and radar receiving filters for generating AN signals as much as possible to Eves. However, the aforementioned scenario is based on the assumption that all user locations are fixed, and legitimate users are highly susceptible to radar interference. This makes it difficult to apply to highly dynamic IoV scenarios.
In addition, many studies have revealed that optimizing the phase-shift matrix of RIS in IRS-enabled ISAC systems can extend their coverage \cite{qin2023joint,salem2022active}.
Accordingly, \emph{Salem et al.} proposed a PLS solution to maximize achievable secrecy rate,
by jointly designing the receiving BF for radar, reflection coefficient matrix of RIS, and BS' transmitting BF \cite{salem2022active}.
Although these results point out that the sensing signals can assist transmission reliability, they do not address the communication security from the perspective of PLS constrained by communication-sensing coupled interference.

\subsection{Radar sensing interference modeling in ISAC-based IoV}

To measure the performance of the sensing capability, in \cite{braun2013co}, \emph{Martin et al.} introduced the concept of \emph{interruption} into radar networks, which refers to the situation where a radar network cannot detect a specific object due to interference from others. A high probability of interference exists in overlapping frequency bands was proved by \emph{Brooker}, then mutual interference between mmWave radar systems operating in 77 GHz and 94 GHz frequency bands was investigated \cite{brooker2007mutual}. Subsequently, many researches have developed mathematical models to make a deeper understanding of the mutual interference among radars and predict the degree of sensing interference in different scenarios. 
For instance, 
in \cite{al2018stochastic}, \emph{Al-Hourani et al.} pioneered the modelling of automotive radar sensing interference based on stochastic geometry tools. They analyzed the interference between opposing lane radars using Poisson and Lattice models and estimated the SRP based on closed-form interference statistics. 
In \cite{Fang2020Stochastic}, \emph{Fang et al.} studied radar sensing interference in bidirectional multilane scenarios and modelled target radar cross section (RCS) fluctuations using Swerling I model and Chi-Square model, obtaining closed-form expressions for radar sensing SRP. These studies provide the foundation for the integration of communication and sensing in IoV systems.

To better understand communication-sensing coexisting networks, in \cite{Zhang2024Coexistence}, considering radar and communication systems, \emph{Zhang et al.} focused on joint design of communication precoders, radar transmit waveforms, and receiving filters to suppress mutual interference between communication and sensing. 
To deal with the communication-sensing coupled interference, in \cite{su2020secure}, \emph{Su et al.} introduced AN scheme at the source to minimize the SINR at sensing radar, thereby ensuring the expected signal strength of legitimate users. However, the transmit power is increased by introducing external AN into the system. To mitigate power cost, \emph{Lynggaard} employed the assumption of perfect CSI to predict the minimum transmit power required to overcome interference, at the cost of an increased computational complexity \cite{lynggaard2018using}. From above works, to some extent, we can find that although the proposed methods have improved the performance of communication and sensing, additional power consumption and computational burden were also introduced.

Table \ref{tab:methods} summaries the existing PLS methods and highlights the differences between our proposed and other works.
When applying the radar sensing interference in PLS design in IoV, the sensing accuracy, transmission reliability and security should be considered simultaneously. Therefore, on the one hand, the performance analytical framework of ISAC systems considering communication-sensing coupled interference should be established, which describes the relationship between communication and sensing capabilities. On other hand, from the perspective of radar sensing interference utilization, no studies investigate the confidential communication by jointly designing transmission power and trajectory of the source.
This motivates our research in this work. 
In addition, the difference between our work and previous studies is described as follows.
\begin{itemize}
  \item Different from PLS methods based on AN schemes \cite{wang2020physical,Zhang2019Transmit,Xu2021Low}, which did not consider the communication-sensing coupled interference, we make an effective transformation from ``communication interference based security'' to ``sensing interference based security'';
  \item In the context of ISAC-IoV, different from using sensing/AN interference in studies of PLS design \cite{salem2022active,su2020secure}, only the sensing interference is utilized to secure wireless communications, without extra energy resources;
  \item Instead of transmitter trajectory optimization \cite{Li2019UAV,Xu2021Low,Sun2021Unmanned}, the duration of confidential information transmission constrained by limited horizontal angle of Carols' sensing beamwidth, the trajectory and transmission power of the Alice are considered simultaneously.
\end{itemize}

\section{Network Model and Preliminaries}\label{sec:network model}
\begin{figure}[!ht]
\centering
\includegraphics[width=2.6in]{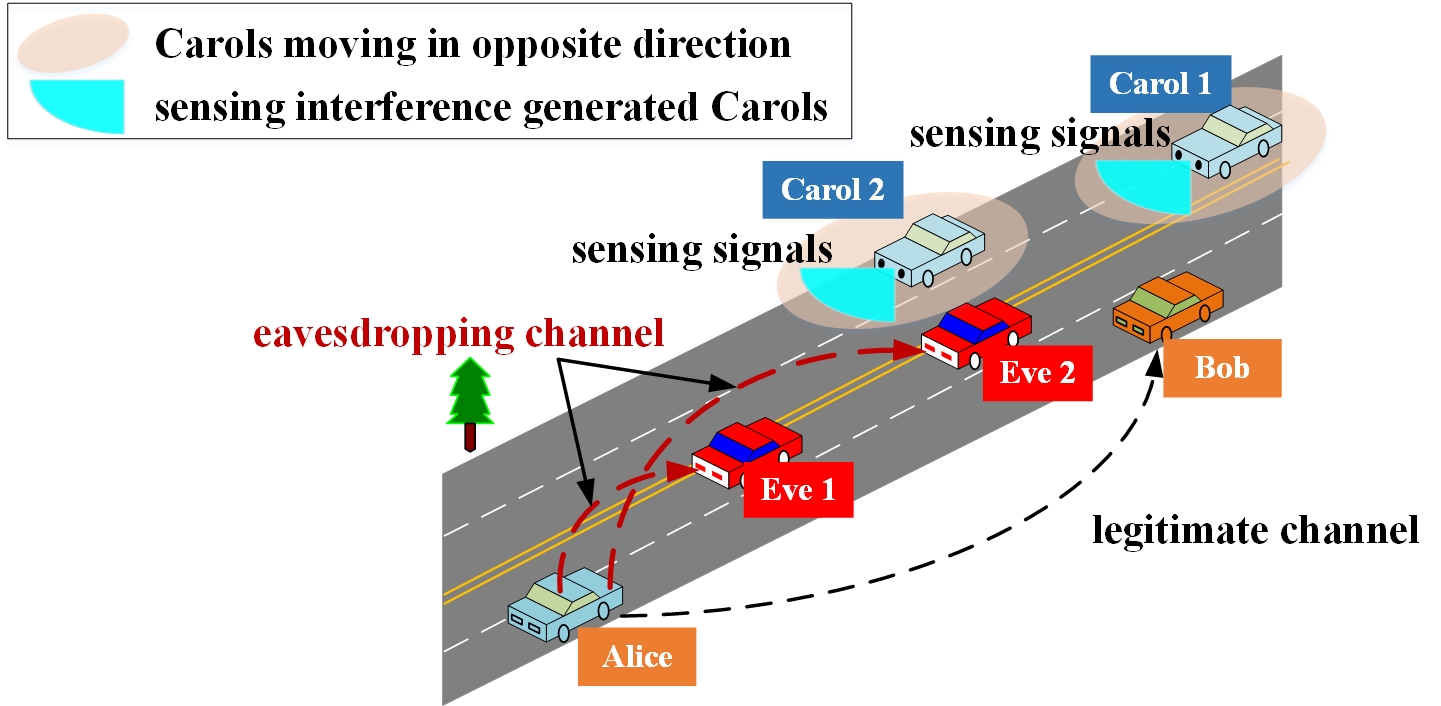}
\caption{\small System model in an ISAC-IoV}
\label{fig:network model}
\end{figure}

To ensure safety driving, each vehicle posses capabilities of sensing and communication working at the same mmWave frequency band. 
In this way, vehicles can sense, monitor and gather the information of a given object or surroundings, and further share the information by communicating with each other. Moreover, there exists coupling-interference between sensing and communications. As shown in Fig. \ref{fig:network model}, given a two-way four-lane scenario, a legitimate source vehicle (called as Alice)
transmits the confidential information to an intended receiving vehicle (called as Bob) within the communication duration of $T$, in the presence of multiple Eves, which try to intercept the transmitted information from the Alice. In particular, the Alice, Bob and Eves drive in the same direction. In addition, other legitimate vehicles, moving in opposite direction with the Alice, Bob and Eves and acting as interferers (called as Carols) since they generate sensing signals by using the forward-sensing radars can decrease the quality of legitimate channel and eavesdropping channel, sense the target ahead by using their forward-sensing radars and have no communication requirement. Without loss of generality, it is assumed that the positions of Eves and Carols follow the independent Poisson Point Process (PPP) with the densities of $\lambda_i$ and $\lambda_e$, and the sets of corresponding positions are denoted by $\Phi_i$ and $\Phi_e$, respectively. 

\begin{figure}[!ht]
\centering
\includegraphics[width=2.2in]{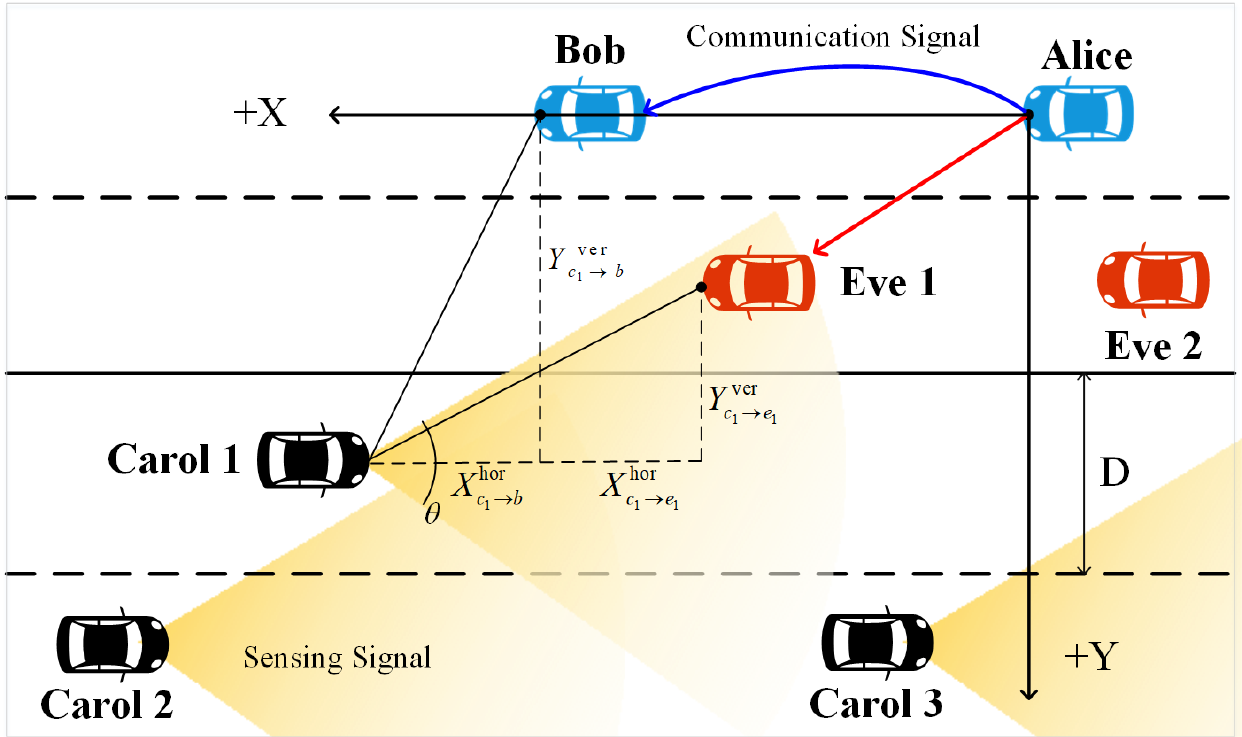}
\caption{\small A simplified version of an ISAC-IoV}
\label{fig:scenario}
\end{figure}

A simplified version of Fig. \ref{fig:network model} is given in Fig.\ref{fig:scenario}, a Cartesian coordinate system can be established with the Alice, Bob, and Eves moving along the positive $x$-axis, while Carols moving along the negative $x$-axis. 
All vehicles are assumed to travel in the center of the road. To ensure the road safety, the width of each lane is denoted by $D_{\rm lane}$ and affordable minimum following distance between any two vehicles is $D^{\min}_{\rm foll}$. The horizontal distance and vertical distance between the Bob and the $i$-th Carol are denoted as $X^{\rm hor}_{c_i\rightarrow b}$ and $Y^{\rm ver}_{c_i\rightarrow b}$, respectively. Similarly, the horizontal distance and vertical distance between the $i$-th Carol and the $j$-th Eve are denoted as $X^{\rm hor}_{c_i\rightarrow e_j}$ and $Y^{\rm ver}_{c_i\rightarrow e_j}$, respectively. 
Due to the limited sensing power, the maximum sensing distance and maximum horizontal angle of sensing beamwidth of the vehicles are limited, which are denoted as $R_{\max}$, and $\theta$, respectively. 

In addition, for better describing physical layer security, we discretize the time duration $T$ into $N$ time slots, indexed by $k$, with the length of $T /N$, under which the transmitting/sensing signal is generated at the begin of the time slot, and received by the vehicles at the end of the current time slot or another one. It is assumed that the Alice transmits the confidential information at each time slot $1 \leq k \leq T /N$. The speed and acceleration at the $k$-th time slot of the $i$-th vehicle are denoted by $v_i[k]$ and $a_i[k]$, and allowable maximization ones are represented as $v_{\max}$ and $a_{\max}$, respectively.

\subsection{Radar sensing model}
As shown in Fig. \ref{fig:scenario}, when the Alice senses and monitors the target Bob ahead by using radar echo, the precision of radar performance is affected by two kinds of interfering signals: 1) the radar echo from non-intended target; 2) radar echos from Carols in the opposite lances. For the sake of simplified analysis, only the latter interference is focused \cite{Fang2020Stochastic,Ghatak2022Radar}\footnote{In radar systems of the vehicles, the radar echoes generated by vehicles in the opposite lanes directly affect the radar system of the current vehicle.
On the one hand, these non-intended echoes are strong and occur frequently because their radar signals are directly aligned with each other. Prioritizing them helps ensure that the model can address the most severe and common interference scenarios.
On the other hand, non-target radar echoes include reflected signals from buildings, road signs, and other environmental objects along the road. These echo signals are usually weak and often present as short, intermittent interference. They can be effectively mitigated through filtering and signal processing techniques. Therefore, it is reasonable to ignore this type of interference to simplify the model and focus on addressing the primary sources of interference.}.
Generally, the RCS of the target is employed to ascertain the extent being detected. Due to the mobility of the vehicles, the radar echo from a target is fluctuating from scan to scan, which is vital to the precision of radar performance. From the conclusion in \cite{lewinski1983nonstationary}, it can be noticed that Chi-square model is a reasonable model to evaluate the fluctuation effectively.
The probability density function (PDF) of degree 2 $\delta$ under Chi-square model can be expressed as
\begin{equation} \label{eq:p_sigma }
p \left( \sigma  \right) =\frac{\delta }{\Gamma \left( \delta \right){\bar{\sigma }_{av}}^{\\delta}}{{\left( \sigma  \right)}^{\delta-1}}\exp \left( -\frac{\delta}{{\bar{\sigma }_{av}}}\sigma  \right),\sigma \ge 0,
\end{equation}
where $\delta$ is the degree of freedom (DoF), and $\bar{\sigma}_{av}$ denotes and average value of the RCS. To simplify the notation, the distance between the vehicle and its sensing target is denoted as $R_{\rm{tar}}$.
Moreover, based on the conclusion in \cite{series2014systems}, for a given RCS of the target, denoted by $\sigma$, the radar echo of a vehicle to the desired target can be written as 
\begin{displaymath}
{P}_{\text{echo}}=\frac{P_{\rm sen}G_t G_r \lambda^{2}_{w}\sigma}{(4\pi)^3{R_{\rm tar}}^{2\alpha}},
\end{displaymath}
where $G_t$ and $G_r$ represent the gains of the radar transmitting antenna and receiving antenna, respectively, $\lambda _{w}$ denotes the wavelength of the sensing signal, $\alpha$ represents the path loss exponent, $P_{\rm sen}$ is the sensing power of the vehicles.

Due to vehicles' mobility, as well as limited sensing distance and horizontal sensing beamwidth of the radars in vehicles, the channel qualities of the Bob and Eves are time-varying. From the perspective of the physical layer security, the Bob does not suffer from the sensing interference generated by the Carols, while the Eves are covered by radar sensing interference as much as possible. Accordingly, on the one hand, the horizontal distance the Bob and the $i$-th Carol should satisfy the condition 
\begin{equation}\label{eq:distance bob without interference}
X^{\rm hor}_{c_i\rightarrow b}> \sqrt{R_{\max}^2 - {Y^{\rm hor}_{c_i\rightarrow b}}{^2}} ~~\text{or}~~ X^{\rm hor}_{c_i\rightarrow b}< \frac{Y^{\rm hor}_{c_i\rightarrow b}}{\tan \theta/2}.
\end{equation}
Otherwise, the Bob will be affected by Carol's sensing signals. On the other hand, to position the Eves within the Carols' sensing range, the horizontal distance between the $i$-th Carol and the $j$-th Eve should fulfil the following condition 
\begin{equation}\label{eq:distance eve being interfered}
\frac{Y^{\rm hor}_{c_i\rightarrow e_j}}{\tan\theta/2}\le X^{\rm hor}_{c_i\rightarrow e_j} \le \sqrt{{{R}_{\max }}^{2}-{Y^{\rm hor}_{c_i\rightarrow e_j}}^{2}}.    
\end{equation}
In addition, considering the sensing accuracy of the Alice, When the Alice is sensing and monitoring the Bob, the condition that it does not suffer from potential sensing interference generated by the Carols should satisfy
\begin{equation}\label{eq:distance alice being interfered}
{{X}^{\rm hor}_{c_i\rightarrow a}}> \sqrt{{{R}_{\max }}^{2}-{{Y}^{\rm ver}_{c_i\rightarrow a}}^{2}} ~~ \text{or} ~~{{X}^{\rm hor}_{c_i\rightarrow a}}< \frac{{{Y}^{\rm ver}_{c_i\rightarrow a}}}{\tan \theta/2}.
\end{equation}

Let ${{\Phi }_{c_i\to b}}=\left\{c_i:c_i\in {{\Phi }_{i}},\mathcal{M}\left( i \right)=1 \right\}$, ${{\Phi }_{c_i\to e_j}}=\left\{c_i:\right.$ $\left.c_i\in {{\Phi }_{i}},\right.$$\left. e_j\in\Phi_e,\mathcal{H}\left( i, j \right)=1 \right\}$, and ${{\Phi }_{c_i\to a}}=\left\{c_i:c_i\in {{\Phi }_{i}},\mathcal{F}\left( i \right)=1 \right\}$ denote the sets of Carols covering the Bob, the $j$-th Eve, and the Alice by using sensing signals, respectively, where $\mathcal{M}\left( i \right)$, $\mathcal{H}\left( i, j \right)$, and $\mathcal{F}\left( i \right)$ are indicator functions and given by 
\begin{displaymath}
\mathcal{M}\left( i \right)= \begin{cases}
  & 1,~~\text{otherwise;} \\ 
 & 0,~~\text{satisfying the Eq. \eqref{eq:distance bob without interference};}\\ 
\end{cases} 
\end{displaymath}
\begin{displaymath}
\mathcal{H}\left( i,j \right)= \begin{cases}
  & 0,~~\text{otherswise;}\\ 
  & 1,~~\text{satisfying Eq. \eqref{eq:distance eve being interfered};}
\end{cases}
\end{displaymath}
and 
\begin{displaymath}
\mathcal{F}\left( i \right)= \begin{cases}
  & 1, ~~\text{otherwise;} \\ 
 & 0, ~~\text{satisfying the Eq. \eqref{eq:distance alice being interfered};}\\ 
\end{cases}
\end{displaymath}
respectively. Based on the results in \cite{Fang2020Stochastic}, the cumulative interference caused by Carols in the opposing lane to the Bob and the $j$-th Eve can be represented as
\begin{displaymath}
{{I}_{\rm Bob}}=\sum\limits_{c_i\in {{\Phi }_{c_i\to b}}}{{{A}_{\rm ea}}\cdot S\cdot h_{c_i\rightarrow b}}\cdot{\Vert(X^{\rm hor}_{c_i\rightarrow b},Y^{\rm ver}_{c_i\rightarrow b})\Vert}^{-\alpha },
\end{displaymath}
\begin{displaymath}
{{I}_{e_j}}=\sum\limits_{c_i\in {{\Phi }_{c_i\to e_j}}}{{{A}_{\rm ea}}\cdot S\cdot h_{c_i\rightarrow e_j}}\cdot{\Vert(X^{\rm hor}_{c_i\rightarrow e_j},Y^{\rm ver}_{c_i\rightarrow e_j})\Vert}^{-\alpha },
\end{displaymath}
and
\begin{displaymath}
{{I}_{\rm Alice}}=\sum\limits_{c_i\in {{\Phi }_{c_i\to a}}}{{{A}_{\rm ea}}\cdot S\cdot h_{c_i\rightarrow a}}\cdot{\Vert(X^{\rm hor}_{c_i\rightarrow a},Y^{\rm ver}_{c_i\rightarrow a})\Vert}^{-\alpha },
\end{displaymath}
where ${{A}_{\rm ea}}=\frac{{{G}_{r}}{{\lambda }_{\omega }}^{2}}{4\pi }$ denotes the effective aperture of the radar receiver, ${S}=\frac{{P}{G_t}}{4\pi }$ is the power density at the unit distance from the interfering source, $h$ is the small scale fading caused by multi-path propagation \cite{Fang2020Stochastic}. The path loss is proportional to
${\Vert(X^{\rm hor}_{c_i\rightarrow b},Y^{\rm ver}_{c_i\rightarrow b})\Vert}^{-\alpha}$, where $\alpha \in [2, 6]$ denotes the path-loss exponent, and $\Vert\cdot\Vert$ refers to the Euclidean distance between the Carol and the the typical vehicle (i.e., Bob, Eve, and Alice).

\subsection{Communication model}
When the Alice transmits the confidential information to the Bob, the latter suffers from the potential sensing interference generated by the Carols. In addition, the Bob also receives its radar echo signal using sensing and monitoring the target. As a result, the received signal-interference-noise ratio (SINR) at the Bob can be expressed as
\begin{equation}\label{eq:sinr Bob}
{ {\gamma}^{\rm com}_{\rm Bob}  }=\frac{P_{\rm com}h_{a\rightarrow b} {\Vert(X^{\rm hor}_{a\rightarrow b},Y^{\rm ver}_{a\rightarrow b})\Vert}^{-\alpha}}{\sigma^{2}_{b}+ P^{\rm Bob}_{\rm echo} + I_{\rm Bob}},
\end{equation}
where $P_{\rm com}$ denotes the transmit power of the Alice, $h_{a\rightarrow b}$ denotes the small scale fading between the Alice and the Bob, $\sigma^{2}_{b}$ is the additive white gaussian noise (AWGN) at the Bob, and $P^{\rm Bob}_{\rm {echo}}$ refers to the radar echo power from the itself.

Similarly, when the $j$-th Eve tries to intercept the transmitted message, the received SINR at the $j$-th Eve can be represented as
\begin{equation}
{ {\gamma}^{\rm com}_{e_j}}=\frac{P_{\rm com}h_{a\rightarrow e_j} {\Vert(X^{\rm hor}_{a\rightarrow e_j},Y^{\rm ver}_{a\rightarrow e_j})\Vert}^{-\alpha}}{\sigma^{2}_{e_j}+ P^{e_j}_{\rm {echo}} + I_{e_j}},
\end{equation}
where $h_{a\rightarrow e_j}$ denotes the small scale fading between the Alice and the $j$-th Eve, $\sigma^{2}_{e_j}$ is the AWGN at the $j$-th Eve, and $P^{e_j}_{\rm {echo}}$ refers to the radar echo power from the itself.

\subsection{Assumption-Effectiveness of the Perfect CSI}
The sensing capability helps acquire the CSI and positions of the vehicles accurately, thereby enhancing the communication performance \cite{yu2023integrated}. Similar channel estimation techniques recently have extended to the studies of IoV \cite{chu2023joint,lynggaard2018using}. Hence, we assume the availability of perfect CSI for the Alice-Bob, during the whole transmission period. On the other hand, for the Eve, to better intercept
the confidentiality information transmitted by Alice, it is more likely that the Carols can sense the Eves by using their forward-sensing radars. Accordingly, information sharing among legitimate vehicles can facilitate channel estimation of the Alice-Eve. Therefore, the acquired CSI is expected to be also perfect. Moreover, apart from forward-sensing radars, the vehicles typically integrate various sensors (e.g., smart cameras and light detection and ranging) \cite{Sengupta2019A}, the perfect CSI of the Bob and Eve can be also obtained.

\subsection{Performance metrics: transmission reliability and security, sensing accuracy, and secrecy rate}
\textbf{The connection outage probability (COP)}: Based on the SINR expression in Eq. \eqref{eq:sinr Bob}, the transmission reliability can be measured by determining the COP that the transmitted information from Alice fails to be received successfully at the Bob. Mathematically, the COP can be described as \cite{Yu2024SuRLLC}
\begin{equation} \label{eq:P_co}
{p_{co}}=1 - \mathrm{Pr}\left[ {\gamma^{\rm com}_{\rm Bob} \ge {\beta _b}} \right],
\end{equation}
where $\beta_b$ denotes a given threshold for decoding confidential information determined by the hardware.

\textbf{The secrecy outage probability (SOP)}: To describe the security level that the Eves intercept the confidential information successfully, the SOP is introduced and refers to that at least an Eve's channel capacity exceeds a given decoding threshold $\beta _e$. Mathematically, the SOP can be represented as
\begin{equation}\label{eq:concept of sop}
{p_{so}} = 1 - \mathrm{Pr}\left[ {\gamma^{\rm com}_{e_j} < {\beta _e}|\text{for all $e_j\in \Phi_e$}} \right].
\end{equation}

\textbf{The success range probability (SRP)}: Similarly, to evaluate the sensing accuracy level of the Alice, the SRP refers to  the probability of successfully detecting an intended target. That is, the SRP at Alice can be expressed as
\begin{equation} \label{eq:SRP0}
{{p}_{sr}}=\mathrm{Pr} \left[ \gamma^{\rm sen}_{\rm Alice}>\beta_s \right],
\end{equation}
where $\beta_s$ is a given sensing threshold of the Alice determined by the hardware, and $\gamma^{\rm sen}_{\rm Alice}$ is represented as
\begin{displaymath} 
{\gamma}^{\rm sen}_{\rm Alice} =\frac{P^{\rm Alice}_{\rm{echo}}}{\sigma^2_a+\sum\limits_{c_i\in {{\Phi }_{c_i\rightarrow a}}}{{{A}_{\rm ea}}Sh_{c_i\rightarrow a}}{\Vert (X^{\rm hor}_{c_i\rightarrow a}, Y^{\rm ver}_{c_i\rightarrow a})\Vert}^{-\alpha }}.
\end{displaymath}

\textbf{The transmission reliability and sensing accuracy based secrecy rate ($\rm TRSA\_SR$) in worst case}: The secrecy rate is an intuitive performance metric that describes channel dominance \cite{Vuppala2018On}. When the secrecy rate is positive, it means that the perfect secrecy can be achieved. In this paper, to simultaneously ensure both the expected performances in terms of communication and sensing, we define the concept of the $\rm TRSA\_SR$ in the worst case, which is represented as
\begin{equation}
\tau[k]  =( 1-p_{co}){{p}_{sr}}\frac{1}{N}\sum\limits_{k=1}^{N} \left[ {R_{\rm Bob}[k]}-\max_{e_j\in\Phi_e}{{R_{e_j}[k]}}\right]^+,
\end{equation}
where $R_{\rm Bob}[k]=\log_2(1 + \gamma^{\rm com}_{\rm Bob})$ represents the channel capacity from the Alice to the Bob, and $R_{e_j}[k]=\log_2(1+\gamma^{\rm com}_{e_j})$ represents the channel capacity from the Alice to the $j$-th Eve at $k$-th time slot, $a=[b-c]^+$ indicates that $a=b-c$ if $a\geq b$, and $a=0$ otherwise. When the secrecy rate is negative, the transmit power is set to zero.

\subsection{Optimization problem formulation }
Due to limited horizontal angle of Carols' radar sensing beamwidth and vehicles' mobility, the Eve only suffers from radar sensing interference during a series of time slots. Therefore, we should first determine the duration of confidential information transmission. Let $k_{\rm start}$ and $k_{\rm{end}}$ denote the indexes of the earliest time slot and the latest time slot for confidential signal transmission at the Alice, respectively. Then, to achieve the communication reliability and security, as well as sensing accuracy in the IoV, we investigate the joint optimization design of the transmit power $\textbf{P}_{\rm com} \triangleq {\left[ P_{\rm com}[k_{\rm start}],\ldots ,P_{\rm com}[k_{\rm end}] \right]}^T $ and straight trajectory along the $x$-axis $\textbf{X}_{\rm Alice} \triangleq {\left[ X_{\rm Alice}[k_{\rm start}],\ldots ,X_{\rm Alice}[k_{\rm end}] \right]}^T$ to maximize the $\rm TRSA\_SR$ of the system over all $N$ time slots. Note that, due to the fact that the Bob and the Eves suffer from the radar sensing interference generated by the Carols. Therefore, before jointly optimizing the transmit power and straight trajectory of the Alice, we should find the optimal potential time duration of information transmission, under which the Eves are covered by the radar sensing interference of the Carols, denoted by $[t_{\rm start}, t_{\rm end}]$, where $1\leq k_{\rm start}\leq k_{\rm end}\leq T/N$. To sum up, we formulated the $\rm TRSA\_SR$ maximization problem as follows.
\begin{subequations} \label{eq:opt0}
\begin{align}
\text{Objective:}~&\max\limits_{\textbf{P}_{\rm com}, \textbf{X}_{\rm Alice}} \tau[k] \\
\mbox{s.t.}\quad
& 0\le P_{\rm com}\left[ k \right]\le {{P}^{\rm com}_{\max }},  
\forall k_{\rm start}\le k\le k_{\rm end}  
  \label{eq:opt0-P},  \\
&|{{v}_{i}}[k] |\le {v_{\max }} \label{eq:opt0-v}, \\
& |{{a}_{i}}[k]|\le a_{\max } \label{eq:opt0-a},\\
& X^{\rm hor}_{a\rightarrow b}[ k] \le D^{\min}_{\rm foll} \label{eq:opt0-x},\\
&{1-{p}_{co}}\ge \zeta _{\min }^{\rm rel} \label{eq:opt0-Pco},\\
&{p}_{so}\le \zeta _{\max }^{\rm sec}\label{eq:opt0-Pso}  ,\\
&{p}_{sr}\ge \zeta _{\min }^{\rm sen}\label{eq:opt0-Psr},
\end{align}
\end{subequations}
where $P^{\rm com}_{\max}$ is the maximum transmission power for confidential information in Eq. \eqref{eq:opt0-P}, constraint Eq. \eqref{eq:opt0-v} restricts vehicle speeds, and $|\cdot|$ is the absolute value since the velocity is a vector and possesses directionality. Next, constraint \eqref{eq:opt0-a} limits vehicle acceleration and constraint \eqref{eq:opt0-x} imposes a safety distance between two vehicles in the same lane for safety. Constraints \eqref{eq:opt0-Pco} and \eqref{eq:opt0-Pso} respectively enforce reliability and security thresholds for communication, denoted by $\zeta _{\min }^{\rm rel}$ and $\zeta _{\max }^{\rm sec}$. Constraint \eqref{eq:opt0-Psr} represents performance requirements for sensing accuracy, where $\zeta _{\min }^{\rm sen}$ is the minimum SRP.
However, problem \eqref{eq:opt0} is non-convex
optimization, which is difficult to be solved directly, due to the following reasons: 1) the operator ${\left[ \cdot \right]}^+$ is non-smoothness for the objective function; 2) the non-convexity of objective function with respect to ($\textbf{X}_{\rm Alice}$, $\textbf{P}_{\rm{com}}$) without the operation of ${\left[ \cdot \right]}^+$; 3) the transmission power $\textbf{P}_{\rm{com}}$ is coupled with $\textbf{X}_{\rm Alice}$, which makes the problem more challenging to be solved. 
Although there is no general approach to solving problem \eqref{eq:opt0} optimally, in the following section \ref{sec:opt}, we propose an algorithm to address problem \eqref{eq:opt0} effectively.

\section{Performance Analysis and Confidential Rate Maximisation}\label{sec:opt}
In this section, first of all, with the help of the tools of stochastic geometry, we establish the closed-form expressions of the COP, SOP, and SRP. Then, the non-convex optimization problem is effectively solved by utilizing methods of Lagrangian multipliers, auxiliary variables, first Taylor expansion and successive convex approximation (SCA). Finally, we conduct a theoretical analysis of the convergence and computational complexity of the proposed algorithm.

\subsection{Analytical framework of closed-form expressions of the COP, SOP, SRP and secrecy rate maximization} 
Based on the PDF of the RCS in Eq. \eqref{eq:p_sigma } and the tools of stochastic geometry, we consider the scenario of $\alpha=4$ across time slots, and derive the closed-form results. This case allows the closed-form results to be derived, because under this case the complex integration involved in the calculation can be solved, and provides a guideline to analyze the transmission reliability and security, as well as the sensing accuracy. The corresponding reasonability for deriving closed-form expression has been discussed in the studies of \cite{ Yu2024SuRLLC,Yu2024A}.\footnote{In urban environments with buildings, vehicles, and other obstacles, the settings of $\alpha=4$ is a reasonable approximation that the signal experiences significant attenuation. In addition, empirical studies often support the use of $\alpha=4$ in certain environments, as it closely matches the observed signal attenuation patterns. Therefore, the assumption of $\alpha=4$ not only simplifies the derivation of closed-form expressions for outage probabilities in PLS analysis, but also accurately represents common urban environments.
}

\begin{theorem}
Given a distance between the Alice and the Bob and RCS of the target sensed by the latter, denoted by $\bar{X}^{\rm hor}_{a\rightarrow b}$ and $\bar{\sigma}_{\rm Bob}$, the closed-form expression of the corresponding COP is given by 
\begin{equation} \label{eq:P_co final}
\begin{aligned}
{{P}_{\text{co}}} &=1-\exp \left[ -\pi {{\lambda }_{i}}{{\left( \frac{{{\beta }_{b}}{{A}_{\rm ea}}S}{P_{\rm {com}}} \right)}^{\frac{1}{2}}}
{{\bar{X}}_{a\rightarrow b} ^{\rm {hor}}} {^{2}}
{\cdot C \left(4 \right)} \right.\\
&\left. ~~~ -\frac{{{\beta }_{b}}}{P_{\rm com}} \left({\sigma^{2}_{b}+ {\frac{P{{G}_{t}}{{G}_{r}}{{\lambda }_{\omega}}^{2}\bar{\sigma}_{\rm Bob} }{{{\left( 4\pi  \right)}^{3}}
{R_{\rm{tar}}}^{8 }}}} \right){\bar{X}^{\rm hor}_{a\rightarrow b}}{^{4}}\right],
\end{aligned}
\end{equation}
where $C \left(\alpha \right)= \Gamma \left( 1+\frac{2}{\alpha } \right) \Gamma \left( 1-\frac{2}{\alpha } \right)$, and $\Gamma(\cdot )$ is the Gamma function. 
\end{theorem}

\begin{IEEEproof}
Due to the limited space of the manuscript, the detail derivation process of Eq. \eqref{eq:P_co final} can refer to \cite{Lkx2024}.
\end{IEEEproof}

\begin{theorem}
For the case of $\alpha=4$ and a given RCS of the target, denoted by $\bar{\sigma}_{e_j}$, sensed by the $j$-th Eve, the upper bound and lower bound of the SOP happening at this Eve are given by Eq. \eqref{eq:sop_u} and Eq. \eqref{eq:SOP_L}, respectively, where $ \mathrm{Erfc} \left( z \right)= { \frac{2}{\sqrt{\pi}}} {\int_{z}^{\infty} { e^{-t^2}}  } dt$ represents the complementary error function. 
\begin{figure*}[t] 
\begin{equation} \label{eq:sop_u}
\begin{aligned}
p_{so}^{\rm upper}= & 1-\exp \left[ -\frac{{{\lambda }_{e}}{{\pi }^{\frac{3}{2}}}}{2\sqrt{\frac{{{\beta }_{e}}}{P_{\rm com}}\left( {\sigma^{2}_{e}+ P^{e_j}_{\rm{echo} }} \right)}}\cdot \exp \left[ \frac{{{\left( \pi {{\lambda }_{i}}C\left( 4 \right) \right)}^{2}}{{A}_{\rm ea}}S}{4 \left( {\sigma^{2}_{e}+ P^{e_j}_{\rm{echo}} } \right) } \right] \cdot \mathrm{Erfc}\left[ \sqrt{\frac{{{A}_{\rm ea}}S}{4 \left( {\sigma^{2}_{e}+ P^{e_j}_{\rm {echo}}} \right)}} \right] \right]
\end{aligned}
\end{equation}
and
\begin{equation} \label{eq:SOP_L}
\begin{aligned}
p_{so}^{\rm lower}= &\exp \left[ -\frac{{{\lambda }_{e}}{{\pi }^{\frac{3}{2}}}}{2\sqrt{\frac{{{\beta }_{e}}}{P_{\rm com}}\left( {\sigma^{2}_{e}+ P^{e_j}_{\rm{echo}}} \right)}} \cdot \exp \left[ \frac{{{\left( \sqrt{\frac{{{\beta }_{e}}{{A}_{\rm ea}}S}{P_{\rm com}}}{{\lambda }_{i}}\pi C\left( 4 \right)+{{\lambda }_{e}} \right)}^{2}}}{4\frac{{{\beta }_{e}}}{P_{\rm com}}\left( {\sigma^{2}_{e}+ P^{e_j}_{\rm{echo}}} \right)} \right] \cdot \mathrm{Erfc}\left[ \frac{\sqrt{\frac{{{\beta }_{e}}{{A}_{\rm ea}}S}{P_{\rm com}}}{{\lambda }_{i}}\pi C\left( 4 \right)+{{\lambda }_{e}}}{2\sqrt{\frac{{{\beta }_{e}}}{P_{\rm com}}\left( {\sigma^{2}_{e}+ P^{e_j}_{\rm{echo}}} \right)}} \right] \right]
\end{aligned}
\end{equation}
where $P^{e_j}_{\rm{echo}} =\frac{P_{\rm sen}G_tG_r{\lambda_{w}}^2 \bar{\sigma}_{e_j}}
{( 4\pi )^3{R_{\rm{tar} }}^{2\alpha }}$.
\hrule 
\end{figure*}
\end{theorem}

\begin{IEEEproof}
Due to the limited space of the manuscript, the detail derivation process of Eq. \eqref{eq:sop_u} and Eq. \eqref{eq:SOP_L} can refer to \cite{Lkx2024}.
\end{IEEEproof}

\begin{theorem}
For the case of $\alpha=4$, the SRP of the Alice can be represented as
\begin{equation}\label{eq: closed expression srp}
\begin{aligned}
  & {{p}_{sr}} =\frac{\Gamma \left( k,\frac{4\pi \delta \beta_s { R_{\rm{tar}}^{8}}}{{\bar{\sigma }_{av}}}\left( \frac{{{\left( 4\pi  \right)}^{2}}}{{\bar{\sigma }_{av}}P_{\rm sen}{{G}_{t}}{{G}_{r}}{{\lambda }_{w}}^{2}}{\sigma^{2}_{a}}+\frac{{{\lambda }_{i}}}{3}{{\left( 2D_{\rm lane} \right)}^{-3}} \right) \right)}{\Gamma \left(\rm{\delta} \right)} \\ 
\end{aligned}
\end{equation}
\end{theorem}

\begin{IEEEproof}
Due to the limited space of the manuscript, the detail derivation process of Eq. \eqref{eq: closed expression srp} can refer to \cite{Lkx2024}.
\end{IEEEproof}

\subsection{Optimizing transmission time and confidentiality rate} 
In this subsection, since the Eves are more closer to the Alice than the Bob, to achieve the highest $\rm TRSA\_SR$, the Alice tries to transmit the confidential information to the Bob only when the Eves are covered by radar sensing signals of the Carols. In this way, we need to determine the earliest transmit time and the latest end time for information transmission between the Alice and the Bob.
Furthermore, we optimize jointly the design of the transmit power and straight trajectory of the Alice to maximize the $\rm TRSA\_SR$. 
To facilitate mathematical calculations, we make an assumption that a single Eve and a single Carol are considered, but the method can be extended to scenarios with multiple Eves and Carols.

\subsubsection{\textbf{Communication duration for information transmission }}
Let $c=\frac{1}{\sqrt{\mu \varepsilon }}$ denote the propagation speeds of communication and sensing signals, where $\mu$ and $\varepsilon$ respectively represent the permittivity and permeability of the medium \cite{Sun2021Unmanned}. Initially, all vehicles move at a constant speed $v$. In the $k$-th time slot, the positions of Alice, Bob, Eve, and Carol are respectively $\left(x_{a}[k],0 \right)$, $\left( x_{b}[k],0\right)$, $\left(  x_{e}[k], D_{\rm{ lane}}\right)$, and $\left(  x_{c}[k], 2D_{\rm{ lane}}\right)$.



\textit{\textbf{Earliest transmission time of the Alice:}}
Let $ k_{c}^{\rm {ear}}$ denote the index of the earliest time slot that the Carol generates radar sensing signal, and $k_{e}^{\rm{ear}}$ denote another index of the earliest time slot that the Eve receives the signal.
Next, we first determine the relationship between $ k_{c}^{\rm {ear}}$ and $ k_{e}^{\rm {ear}}$.
For this case, the maximum horizontal distance between the Carol and the Eve is $X^{\max, \rm{hor}}_{c\rightarrow e}=\sqrt{{{R}_{\max }}^{2}-{{D_{\rm lane}}^{2}}}$, and
the maximum propagation time of the radar sensing signal reaching to the Eve is $ t^{\max,\rm{sen}}_{{c\rightarrow e}} = R_{\max}/c$. Therefore, the sum of the moving distance of the Eve and the $X^{\max, \rm{hor}}_{c\rightarrow e}$ equals the horizontal distance between the Carol and the Eve at the $k_{c}^{\rm ear}$-th time slot.

\begin{displaymath}
X^{\max,\rm{hor}}_{c\rightarrow e}+ v t^{\max ,\rm{sen} }_{c\rightarrow e} = {x_c}\left[ k_c^{\rm ear} \right]-{x_e}\left[ k_c^{\rm {ear}} \right],
\end{displaymath}
where ${x_c}\left[ k_c^{\rm ear} \right]$ and 
${x_e   }\left[ k_c^{\rm ear} \right]$ represent the positions of Carol and Eve at the $k_c^{\rm ear}$-th time slot, respectively. The needed time that the Eve receives radar sensing signal equals to the sum of the time that Carol sends the radar sensing signal and the radar sensing signal propagation time, $ t \left[ {k_{e}^{\rm ear}} \right]= t^{\max ,\rm{sen}}_{c\rightarrow e} + t \left[{k_c^{\rm ear}} \right]$. The values of $k_c^{\rm ear}$ and $k_e^{\rm ear}$ are
\begin{displaymath}
{k_c^{\rm{ear}}}=\frac{{{x}_{e}\left[ 0 \right] }-{{x}_{c}\left[ 0 \right] }-X^{\max ,\rm{hor}}_{c\rightarrow e } - vt^{\max,\rm{sen}}_{c\rightarrow e}}{2v},
\end{displaymath}
and
\begin{displaymath}
{k_e^{\rm{ear}}}=\frac{{{x}_{e}\left[ 0 \right] }-{x}_{c}\left[ 0 \right] -X^{\max ,\rm{hor}}_{c\rightarrow e }+vt^{\max,\rm{sen}}_{c\rightarrow e}}{2v},
\end{displaymath}
respectively.
Let $k_a^{\rm {ear}}$ be the index of the earliest time slot that the Alice starts to transmit confidential message, and $t^{\rm {com}}_{{a\rightarrow e}}$ be the needed time that the Eve receives the message. Since the Alice and Eve are moving at a constant speed initially, thus $t^{\rm {com}}_{{a\rightarrow e}}$ can be given by
\begin{displaymath}
{{\left( vt^{\rm {com}}_{{a\rightarrow e}}+{{x}_{e} \left[ 0 \right]} \right)}^{2}}+{{D_{\rm lane}}^{2}}={{\left( ct^{\rm {com}}_{{a\rightarrow e}}\right)}^{2}}.
\end{displaymath}
Then, we can get 
\begin{displaymath}
t^{\rm{com}}_{{a\rightarrow e}}=\frac{vx_e \left[ 0 \right] +\sqrt{{{c}^{2}}{{v}^{2}}+{{c}^{2}}{{D_{\rm lane}}^{2}}-{{v}^{2}}{{D_{\rm lane}}^{2}}}}{ \left(  {{c}^{2}}-{{v}^{2}} \right)}.
\end{displaymath}
Moreover, the begin of time slot that the Alice starts to transmit the confidential information equals to the difference between the time that the Eve receives the sensing signal and that of the confidential signal, i.e., $k_a^{\rm {ear}} =  k_e^{\rm {ear}}- t_{a\rightarrow e}^{\rm {com}}  $. Therefore, the index of the earliest time slot that the Alice starts to transmit can be calculated as
\begin{displaymath}
\begin{aligned}
    k_a^{\rm {ear}} = &\frac{{{x}_{e}\left[ 0 \right]}-{{x}_{c}\left[ 0 \right]}-\sqrt{{{R}_{\max }}^{2}-{{D_{\rm lane}}^{2}}}+v\frac{{{R}_{\max }}}{c}}{2v} \\
&-\frac{vx_e \left[ 0 \right]+\sqrt{{{c}^{2}}{{v}^{2}}+{{c}^{2}}{{D_{\rm lane}}^{2}}-{{v}^{2}}{{D_{\rm lane}}^{2}}}}{\left( {{c}^{2}}-{{v}^{2}} \right)}.
\end{aligned}
\end{displaymath}
Recall that $k_{\rm start}$ denote the index of the earliest time slot for the Alice transmitting message.
To ensure the physical layer security potentially, it is expected that the Eve should be covered by sensing signal from the Carol before receiving the confidential message from the Alice, namely $k_{\rm start}\geq k_a^{\rm {ear}}$. 

\textit{\textbf{Latest transmission time of the Alice:}}
Let $k_{c}^{\rm {last}} $ be the index of the index of the latest time slot that the Carol generates the radar sensing signal, and $k_{e}^{\rm {last}}$ denote another index of the lasted time slot when the Eve receives the signal. Next, we first determine the relationship between $k_{c}^{\rm {last}} $ and $k_{e}^{\rm {last}}$. For this case, the minimum horizontal distance between the Carol and the Eve is $X^{\min,\rm{hor} }_{c\to e}=\frac{D_{\rm{ lane}}}{\tan \frac{\theta }{2}}$, and the distance between the Bob and the Carol is then given by $D_{\min }^{c\to e}= \sqrt{X{{^{\min, \rm{hor}}_{c\to e}}^{2}}+{{D_{\rm{ lane}}}^{2}}}$. The minimum propagation time of the radar sensing signal reaching to the Eve is $ t_{c\rightarrow e}^{\min ,\rm{sen}} = D_{\min }^{c\to e}/c$. Therefore, the sum of the moving distance of the Eve and the $X^{\min,\rm{hor} }_{c\to e}=\frac{D_{\rm{ lane}}}{\tan \frac{\theta }{2}}$ equals the horizontal distance between the Carol and the Eve at the $k_{c}^{\rm {ear}}$-th time slot.
\begin{displaymath}
X^{\min ,\rm{hor}}_{c\rightarrow e}+ v t_{c\rightarrow e}^{\min ,\rm{sen}} = {x_c}\left[ k_c^{\rm {last}} \right]-{x_e}\left[ k_c^{\rm {last}} \right],
\end{displaymath}
where ${x_c}\left[ k_c^{\rm {last}} \right]$ and ${x_c}\left[ k_e^{\rm {last}} \right]$ represent the positions of Carol and Eve at the $k_e^{\rm {last}}$-th time slot, respectively. The needed time that
the Eve receives the radar sensing signal equals to the sum of the time that Carol sends the radar sensing signal and the radar sensing signal
propagation time, $ { k_{e}^{\rm {last}} } = {k_c^{\rm {last}}}  +t_{c\rightarrow e}^{\min ,\rm{sen}}$. The values of $k_c^{\rm {last}}$ and $k_e^{\rm {last}}$ are
\begin{displaymath}
{k_c^{\rm {last}}}=\frac{{{x}_{e}\left[ 0 \right] }-{{x}_{c}\left[ 0 \right] }-X^{\min ,\rm{hor}}_{c\rightarrow e} - vt^{\min ,\rm{sen} }_{c\rightarrow e}}{2v},
\end{displaymath}
and
\begin{displaymath}
{k_e^{\rm {last}}}=\frac{{{x}_{e}\left[ 0 \right] }-{{x}_{c}\left[ 0 \right] }-X^{\min ,\rm{hor}}_{c\rightarrow e} + vt^{\min ,\rm{sen} }_{c\rightarrow e}}{2v},
\end{displaymath}
respectively. Moreover, since the time required for the communication signal to be transmitted and received remains constant, $ k_a^{\rm {last}}=  k_e^{\rm {last}} - t_{a\rightarrow e}^{\rm {com}}  $, then the last time when Alice sends the signal can be calculated
\begin{displaymath}
\begin{aligned}
    k_a^{\rm {last}} = &\frac{{{x}_{e}\left[ 0 \right]}-{{x}_{c}\left[ 0 \right]}+\frac{D_{\rm{ lane}}}{\tan \frac{\theta }{2}}\left( \frac{\sec \frac{\theta }{2}}{c}-1 \right)}{2v} \\
&-\frac{vx_e \left[ 0 \right]+\sqrt{{{c}^{2}}{{v}^{2}}+{{c}^{2}}{{D_{\rm lane}}^{2}}-{{v}^{2}}{{D_{\rm lane}}^{2}}}}{\left( {{c}^{2}}-{{v}^{2}} \right)}.
\end{aligned}
\end{displaymath}
Recall that $k_{\rm{end}}$ denote the index of the latest time slot of the Alice transmitting message.  To ensure the PLS potentially, it is expected that the Eve should be covered by radar sensing signal from the Carol before receiving the confidential message from the Alice, namely $k_{\rm{end}} \le  k_a^{\rm {last}} $.


\subsubsection{ \textbf{Joint optimization of transmission power and straight trajectory of the Alice}} 
By jointly optimizing transmission power of the Alice $\textbf{P}_{\rm {com}} \triangleq {\left[ P_{\rm {com}}[k_{\rm start}],\ldots ,P_{\rm {com}}[k_{\rm end}] \right]}^T $ and its straight trajectory along the $x$-axis $\textbf{X}_{\rm Alice} \triangleq {\left[ X_{\rm Alice}[k_{\rm start}],\ldots ,X_{\rm Alice}[k_{\rm end}] \right]}^T$, we aim to maximize the $\rm TRSA\_SR$ over all time slots
under effective communication duration from the $k_{\rm start}$-th time slot to the $k_{\rm end}$-th time slot.
With the settings of $\alpha=2$ \footnote{At the beginning of Section V, we have made an assumption of the number of Alice, Bob, Carol and Eve is one. Thus, LoS channel between Alice and Bob/Eve exists most likely. In addition, all vehicles are going in a straight line. It's a reasonable approximation of signal attenuation mode under LoS channel by setting $\alpha=2$. Furthermore, this configuration provides a baseline for comparisons, enabling an understanding of the fundamental performance limits under ideal conditions before considering other factors. Subsequently, it allows for the study of how additional environmental factors (such as noise power) affects the system performance.}, the problem \eqref{eq:opt0} is formulated as Eq. \eqref{eq:opt1}.
\begin{figure*}[t] 
\begin{equation} \label{eq:opt1}
\begin{aligned}
\text{Objective:}~~&\max\limits_{\textbf{P}_{\rm {com}}, ~\textbf{X}_{\rm Alice}} 
\sum\limits_{k={{t}_{\rm{start}}}}^{{{t}_{\rm{end}}}}\left[ {{\log}_{2}}\left( 1+\frac{{P_{\rm {com}}[k] }{{h}_{a\rightarrow b}}}{( \sigma^{2}_{b}+{{I}_{\rm Bob}} ){{\Vert X^{\rm{hor}}_{a\rightarrow b}\Vert}^{2}}} \right)  -{{\log }_{2}}\left( 1+\frac{{P_{\rm {com}}\left[k\right] }{{h}_{a\rightarrow e}}}{\left(\sigma^{2}_{e}+{{I}_{e}} \right){\Vert(X^{\rm{hor}}_{a\rightarrow e},Y^{\rm ver}_{a\rightarrow e})\Vert}^{2} } \right) \right]\\
\mbox{s.t.}\quad
& \eqref{eq:opt0-P},\eqref{eq:opt0-v},\eqref{eq:opt0-a},\eqref{eq:opt0-Pco},\eqref{eq:opt0-Pso},\eqref{eq:opt0-Psr},\eqref{eq:opt-s.t.x_a}.
\end{aligned}
\end{equation}
\hrule 
\end{figure*}
which is also non-convex and difficult to be solved directly.

In the following, we decompose the problem \eqref{eq:opt1} into two sub-problems for obtaining feasible solutions via the BCD method, namely consisting of transmission power optimization and straight trajectory optimization. In particular, on the one hand, sub-problem (P1) is to optimize the transmit power for a given positions of the Alice. On the other hand, sub-problem (P2) is to optimize the straight trajectory of the Alice for a given transmit power. Finally, by alternately optimizing these two sub-problems to obtain a locally optimal solution, the $\rm TRSA\_SR$ maximization can be obtained.

\textit{\textbf{Transmission power optimization of the Alice:}} Let ${s_1 \left[ k \right]}=\frac{{{h}_{a\rightarrow b}}}{\left( \sigma^{2}_{b}+{{I}_{\rm Bob}} \right){{\Vert X^{\rm{hor}}_{a\rightarrow b} \Vert}^{2}}}$ and ${s_2 \left[ k \right]}=\frac{{{h}_{a\rightarrow e}}}{\left( \sigma^{2}_{e}+{{I}_{e}} \right){\Vert(X^{\rm{hor}}_{a\rightarrow e},Y^{\rm ver}_{a\rightarrow e})\Vert}^{2}}$.
Given a straight trajectory $\bar{\textbf{X}}_{\rm Alice}$ of the Alice, the problem \eqref{eq:opt1} can be simplified into
\begin{subequations} \label{eq:P1}
\begin{align}
\mathrm{P1:}~~& \max\limits_{\textbf{P}_{\rm {com}}} 
\sum\limits_{k=k_{\rm start}}^{k_{\rm end}}\left[ \log_{2}\left( 1+{s_1[k]}P_{\rm {com}}[k]\right)\right.\nonumber\\
&~~~~~~\left.-{{\log }_{2}}\left( 1+{s_2 \left[ k \right]}{P_{\rm {com}}\left[k\right] } \right)\right]\\
\mbox{s.t.}~
& \eqref{eq:opt0-P}.
\end{align}
\end{subequations}
The non-negative weighted sum still retains concavity because the $\rm TRSA\_SR$ is positive and the  objective is a concave function. Therefore, the Lagrangian Multiplier method can be used for the solution \cite{Gopala2008On}, 
namely
\begin{displaymath}
\frac{{{s}_{1}[k]}}{1+{{s}_{1}[k]}P_{\rm {com}}\left[ k \right]}-\frac{{{s}_{2}[k]}}{1+{{s}_{2}[k]}P_{\rm {com}}\left[ k \right]}-\lambda_{\rm lag}=0,
\end{displaymath}
where $\lambda_{\rm lag}$ denotes the Lagrange multiplier, which is obtained from the result of the last iteration. Accordingly, the solution for $P_{\rm {com}}\left[ k \right]$ is
\begin{equation} \label{eq:P-opt}
\begin{aligned} 
P_{\rm {com}}\left[ k \right] &=\frac{1}{2}\left[ \sqrt{{{\left( \frac{1}{{{s}_{1}}\left[ k \right]}-\frac{1}{{{s}_{2}}\left[ k \right]} \right)}^{2}}+\frac{4}{\lambda _{lag}}\left( \frac{1}{{{s}_{2}}\left[ k \right]}-\frac{1}{{{s}_{1}}\left[ k \right]} \right)} \right. \\
& ~~~\left. -\left( \frac{1}{{{s}_{1}}\left[ k \right]}+\frac{1}{{{s}_{2}}\left[ k \right]} \right) \right]. 
\end{aligned}
\end{equation}
Therefore, the optimal solution of the transmit power for problem \eqref{eq:P1} in the $k$-th time slot is represented as
\begin{equation}
{{P}^{*}_{\rm com}}\left[ k \right]=\min \left\{ P_{\rm com}\left[ k \right],{{P}^{\rm com}_{\max }} \right\}.
\end{equation}
Finally, $\textbf{P}^{*}_{\rm com}=\{{{P}^{*}_{\rm com}}[k_{\rm start}], ..., {{P}^{*}_{\rm com}}[k], ..., {{P}^{*}_{\rm com}}[k_{\rm end}]\}$. It is worth noting that when the channel quality for the Eve is better than that of the Bob, the transmit power is set to 0.

\textit{\textbf{Straight trajectory optimization of the Alice:}} For a given transmit power allocation $\bar{\textbf{P}}_{\rm com}$, the objective function is also non-concave by optimizing the driving trajectory of the Alice, rendering it impossible to directly be solved. Therefore, it is necessary to transform the function into a concave one. To facilitate the solution, relaxation variables $u\triangleq {\left[ u[k_{\rm start}],\ldots ,u[k_{\rm end}] \right]}^T$ and $w\triangleq {\left[ w[k_{\rm start}],\ldots ,w[k_{\rm end}] \right]}^T$  are introduced
, and the problem can be rewritten as \cite{Yu2024SuRLLC}
\begin{subequations} \label{eq:P2}
\begin{align}
\mathrm{P2:} \max\limits_{\textbf{X}_{\rm Alice}} & 
\sum\limits_{k={k_{\rm{start}}}}^{{k_{\rm{end}}}}{\left[ {{\log }_{2}}\left( 1+ \frac{s_3\left[k\right]}{u\left[k\right]} \right) -{{\log }_{2}}\left( 1+ \frac{s_4\left[k\right]}{w\left[k\right]} \right)\right]}\\
\mbox{s.t.}\quad
&w\left[ k \right]-x_a{{\left[ k \right]}^{2}}+2{{x}_{e}}\left[ k \right]x_a\left[ k \right]\nonumber \\
&~~~~-{{x}_{e}}{{\left[ k \right]}^{2}} - D_{\rm{lane}}^2 \le 0,~\forall k, \label{eq:s.t.w}\\ 
 & x_a{{\left[ k \right]}^{2}}-2{{x}_{b}}\left[ k \right]x_a\left[ k \right]+{{x}_{b}}{{\left[ k \right]}^{2}}\nonumber\\
 &~~~~-u\left[ k \right]\le 0,~\forall k,  \label{eq:s.t.u}\\ 
& \eqref{eq:opt0-v},\eqref{eq:opt0-a},\eqref{eq:opt0-x},\eqref{eq:opt0-Pco},\eqref{eq:opt0-Pso},\eqref{eq:opt0-Psr}.\nonumber 
\end{align}
\end{subequations}
where ${{s}_{3}}\left[ k \right]=\frac{P_{\rm com}\left[ k \right]{{h}_{a\rightarrow b}}}{\sigma^{2}_{b}+{{I}_{\rm Bob}}}$ and ${{s}_{4}}\left[ k \right]=\frac{P_{\rm com}\left[ k \right]{{h}_{a\rightarrow e}}}{\sigma^{2}_{e}+{{I}_{e}}}$.
It can be shown that the function ${{\log }_{2}}\left( 1+\frac{{{s}_{3}}\left[ k \right]}{u\left[ k \right]} \right)$ is convex with respect to $u\left[ k \right]$, and $-{x_a\left[ k \right]}^2$ is concave with respect to $x_a\left[ k \right]$. Consequently, the Taylor expansion of a convex function provides a lower bound for the original function, while the Taylor expansion of a concave function provides an upper bound. We propose an iterative algorithm to solve problem approximately by applying the SCA method. The algorithm obtains an approximate solution to problem \eqref{eq:P2} by maximizing a concave lower bound of its objective function within a convex feasible region, which is detailed as follows. Given the initial points $\left( x_0,u_0 \right)$, where $x_0\triangleq {\left[ x_0[k_{\rm start}],\ldots ,x_0[k_{\rm end}] \right]}^T$ the original function is approximated as
\begin{equation}
\begin{aligned}
{{\log }_{2}}\left( 1+\frac{{{s}_{3}}\left[ k \right]}{u\left[ k \right]} \right) 
&\ge {{\log }_{2}}\left( 1+\frac{{{s}_{3}}\left[ k \right]}{{{u}_{0}}\left[ k \right]} \right) \\
&-\frac{{{s}_{3}}\left[ k \right]\left( u\left[ k \right]-{{u}_{0}}\left[ k \right] \right)}{\ln 2\left( u_{0}^{2}\left[ k \right]+{{s}_{3}}\left[ k \right]{{u}_{0}}\left[ k \right] \right)} ,\label{eq:opt-log}
\end{aligned}
\end{equation}
and $-x_a^2[k]$ can be represented as
\begin{equation}
-{{x_a}^{2}}\left[ k \right]\le x_{0}^{2}\left[ k \right]-2{{x}_{0}}\left[ k \right]x_a\left[ k \right] \label{eq:opt-s.t.x_a}.
\end{equation}
Neglecting the constant term, the optimization problem can be reduced to
\begin{subequations} \label{eq:P2*}
\begin{align}
\mathrm{P2':} \max\limits_{\textbf{X}_{\rm Alice}} & 
\sum\limits_{k={k_{\rm{start}}}}^{{k_{\rm{end}}}}{\left[  
-\frac{{{s}_{3}}\left[ k \right] u\left[ k \right]}{\ln 2\left( u_{0}^{2}\left[ k \right]+{{s}_{4}}\left[ k \right]{{u}_{0}}\left[ k \right] \right)} \right.} \nonumber \\
&{\left.~~~~~ -{{\log }_{2}}\left( 1+ \frac{s_4\left[k\right]}{v\left[k\right]} \right)\right]} \\
\mbox{s.t.}\quad
&w\left[ k \right]+ x_0{{\left[ k \right]}^{2}} -2x_0[k]x_a[k] +2{{x}_{e}}\left[ k \right]x_a\left[ k \right] \nonumber \\
&-{{x}_{e}}{{\left[ k \right]}^{2}} -D_{\rm{lane}}^2 \label{eq:s.t.u2} \le 0,\\ 
& \eqref{eq:opt0-v},\eqref{eq:opt0-a},\eqref{eq:opt0-x},\eqref{eq:opt0-Pco},\eqref{eq:opt0-Pso},\eqref{eq:opt0-Psr}, \eqref{eq:s.t.u}.\nonumber 
\end{align}
\end{subequations}
The final objective function obtained is concave,
which can be effectively solved using various existing algorithms and toolboxes, such as CVX and interior-point methods. Due to the introduction of slack variables and the use of first-order Taylor method in \eqref{eq:opt-log} to approximate the upper bound of $\rm TRSA\_SR$, the optimal solution of problem \eqref{eq:P2*} can only serve as the lower bound of original question.

\subsection{AO algorithm for $\rm TRSA\_SR$ maximization}
By using AO algorithm to jointly optimize the transmit power and straight trajectory of the Alice, the $\rm TRSA\_SR$ maximization can ultimately be achieved. By alternately optimizing subproblems, the complexity of the problem is effectively reduced, and an approximate solution is obtained through a gradual convergence approach. A tight approximate solution can be obtained by solving the problem \eqref{eq:opt0}.

\begin{algorithm}[H] 
	\caption{Proposed Algorithm for Problem \eqref{eq:opt0}}
	\begin{algorithmic}[1]  
		\State  \textbf{Initialization:} the  maximum iteration number $I_{\max}$, initial solutions of transmit power $P^{0}_{\rm com}$, Alice's position $X^{0}_{\rm Alice}$, $\tau_0 = f_{\eqref{eq:opt0}}(P^{0}_{\rm com}, X^{0}_{\rm Alice})$, index of initial iteration $i=0$, and convergence threshold $\tau_{\varepsilon}$\\
		\textbf{Repeat}\\
		~~~~$i = i + 1$\\
		~~~~Update $P^{i}_{\rm com}$ with initial point $\left({P^{i-1}_{\rm com}}, {X^{ i-1 }_{\rm Alice}} \right)$ with Eq. \eqref{eq:P-opt}\\
		~~~~Update $X^{i}_{\rm Alice}$ with initial point $\left({P^{i}_{\rm com}}, X^{i-1}_{\rm Alice} \right)$ with Eq. $\eqref{eq:P2*}$\\
		~~~~$\tau^i=f_{\eqref{eq:opt0}}({P^{i}_{\rm com},X^{i}_{\rm Alice}})$.\\
		\textbf{Until} $\tau^{i} -\tau^{i-1}< \tau_{\varepsilon}$ or the maximum iteration number $I_{\max}$ is reached
	\end{algorithmic}
	\label{al:HSVM}
\end{algorithm}

\subsubsection{Convergence analysis}
 The optimization algorithm, based on alternating iterations in Algorithm \ref{al:HSVM}, defines the solution of the $\left( i-1 \right)$-th iteration $\left( P^{i-1}_{\rm com}, X^{i-1}_{\rm Alice}\right)$, and the objective function 
$\max\limits_{P^{i-1}_{\rm {com}}, X^{i-1}_{\rm{Alice}
}}\tau^{i-1}$.
In the $i$-th iteration, given $X^{i-1}_{\rm{Alice}}$, the objective can be obtained by solving sub-problem $\mathrm{P1}$, which yields $X^{i}_{\rm{Alice}}$ and we can get $\max\limits_{P^{i}_{\rm com},X^{i-1}_{\rm{Alice}}}\tau^{i-1} \ge \max\limits_{P^{i-1}_{\rm com}, X^{i-1}_{\rm{Alice}}}\tau^{i-1}$.
Similarly, sub-problem $P2'$ can output $X^{i}_{\rm{Alice}}$. Furthermore, $\max\limits_{ P^{i}_{\rm com},X^{i}_{\rm{Alice}}}\tau^{i} \ge \max\limits_{P^{i}_{\rm com}, X^{i-1}_{\rm{Alice}}}\tau^{i-1}$ holds.
It can be demonstrated that the objective function value is non-decreasing following each iteration. Due to the constraints on the power values, the objective function has an upper bound, which proves the convergence of the algorithm.

\subsubsection{complexity analysis} The problem \eqref{eq:opt0} is divided into two subproblems using the BCD method, and the trajectory optimization problem can be solved by using the CVX toolbox, which automatically transforms optimization problems into standard form and solves them using interior-point methods. 
Based on the conclusion that when one of block problems decomposed by BCD algorithm is solved by SCA algorithm, the corresponding computational complexity is the same as that of SCA algorithm for solving a single problem \cite{He2023full}, and results in \cite{Sun2021Unmanned}, the computational complexity of interior point method in each iteration is $O(N^3\log(\epsilon^{-1})$, where $N$ is the number of variables, and $\epsilon$ is the iteration accuracy. To sum up, the complexity of Algorithm \ref{al:HSVM} is $O\left( I_{\rm{iter}} N^3\log(\epsilon^{-1}) \right)$.

\section{Simulation Results}\label{sec:simulation}
In the section, considering the sensing interference empowered PLS, we provide numerical simulations to evaluate the performance of our proposed joint design of straight trajectory and transmit power of the Alice (denoted as \textbf{SI-PLS with ST-TP}). The vehicles moves in the center of the road with the maximum speed $v_{\max}=20 $m/s and maximum acceleration $a_{\max}=3 $m/$\rm s^2$, and affordable minimum following distance between two vehicles $D^{\min}_{\rm foll}=5$m. The maximum transmission power and sensing power are 50dBm and 50dBm, respectively. The maximum iteration number is 30. For comparison, we also consider traditional PLS scheme with no extra power consumption (denoted as \textbf{traditional PLS with no PC}, such as \cite{yin2021uav,jin2024enhanced,Li2024Joint}). Some key parameters and their values are shown in Table \ref{tab:sim para validation}.

\begin{table}[!htb]  \caption{Simulated parameters and values}
\centering
\label{tab:sim para validation}
\begin{tabular}{lll}
\toprule
  Symbol & Meanings &Values\\
\midrule
  $\alpha$            & path-loss exponent                            & [2, 4]         \\
  $R_{\max}$           & maximum sensing distance                    & 200 m \textsuperscript{\cite{series2014systems}}\\
  $F$                 &  frequency band of sensing radar                & 77 GHz         \\
  $\lambda_w$    & wavelength of radar detection signal                                  & 0.0039 m     \\
  $G_t (G_r)$                 & transmitting/receiving antenna gain                                  & 45 dBi \textsuperscript{\cite{al2018stochastic}} \\
  
  $P_{\rm{com}}^{\max}$           & maximum communication power       & 50 dBm    \\
  $P_{\rm{sen}}^{\max}$           & maximum sensing power       & 50 dBm \textsuperscript{\cite{series2014systems} }   \\
  $P_{\rm{com}}$                  & communication power               & 10 dBm    \\
  $P_{\rm sen}$                & sensing power of vehicle's radar      & 10 dBm \textsuperscript{\cite{al2018stochastic}}   \\
  $\theta$             &radar horizontal angle                   & ${60}^\circ $       \\
  $a_0^{\rm{max}}$         & maximum acceleration of the vehicles               & 3 $ \mathrm{m/s^2}$ \\
  $v_{\rm{max}}$           & maximum speed of the vehicles                  & 20 m/s       \\
  $v_{av}$            & average speed of the vehicles                         & 16 m/s       \\
  $D_{\rm lane}$                 & road width                                    & 3.6 m \textsuperscript{\cite{al2018stochastic}}      \\
  $\lambda_I$    & the density of Carols & 0.001 \textsuperscript{\cite{Yu2023The}} \\
  \bottomrule
\end{tabular}
\end{table}

\subsection{Theoretical evaluations of the COP, SOP and SRP}
\begin{figure*}
    \centering
\begin{minipage}{0.24\textwidth}
  \centering
  \includegraphics[width=1.55in]{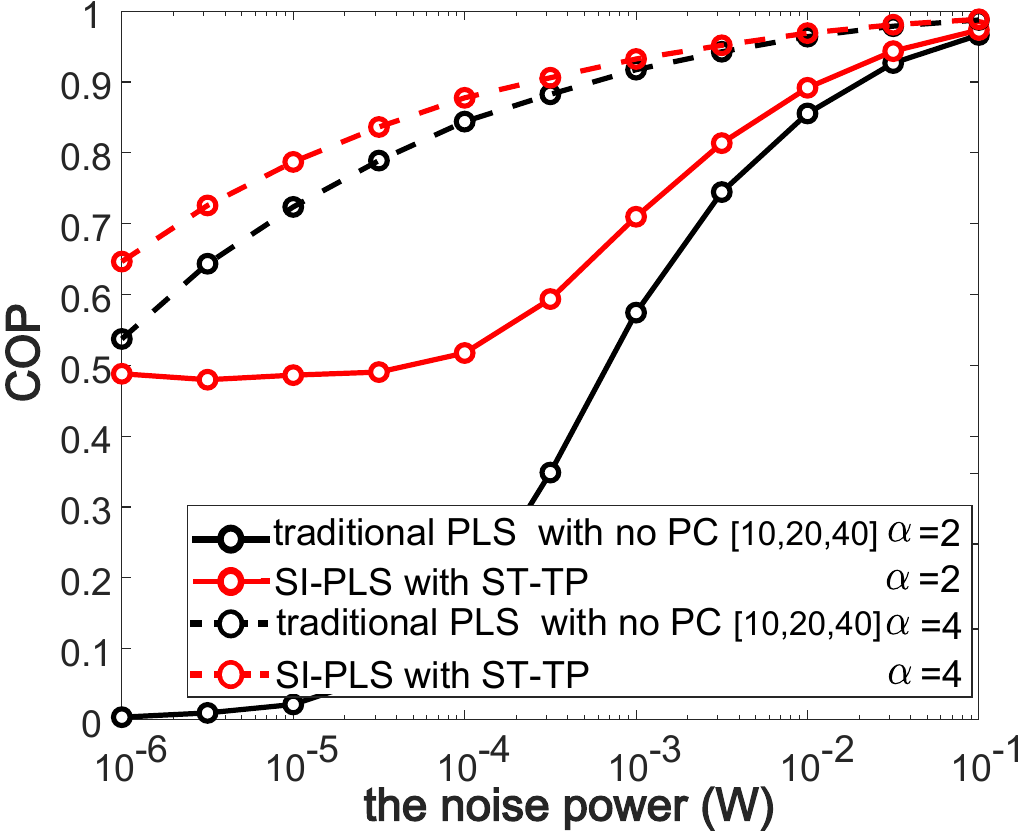}
  \caption{\small COP vs. the noise power}
  \label{fig:COP-Noise}
\end{minipage}
    \hfill
\begin{minipage}{0.24\textwidth}
   \centering
  \includegraphics[height=1.3in,width=1.55in]{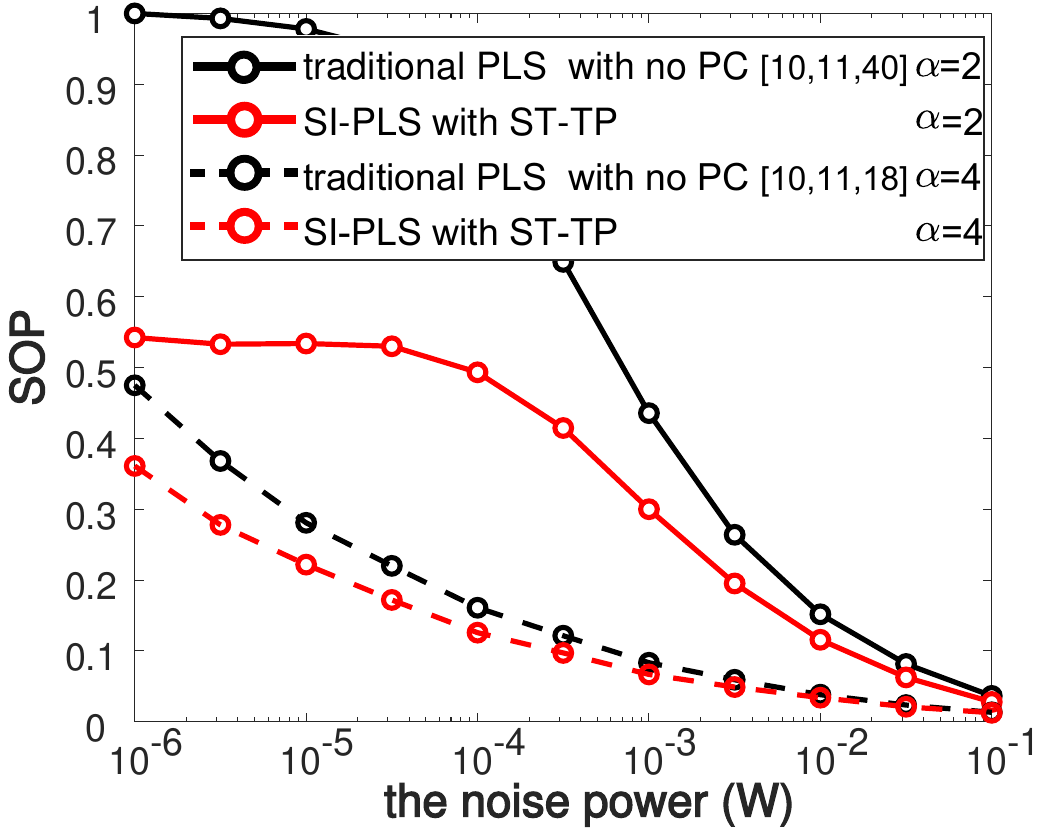}
  \caption{\small SOP vs. noise power}
  \label{fig:SOP-Noise}
\end{minipage}
    \hfill
\begin{minipage}{0.24\textwidth}
        \centering
  \includegraphics[width=1.55in]{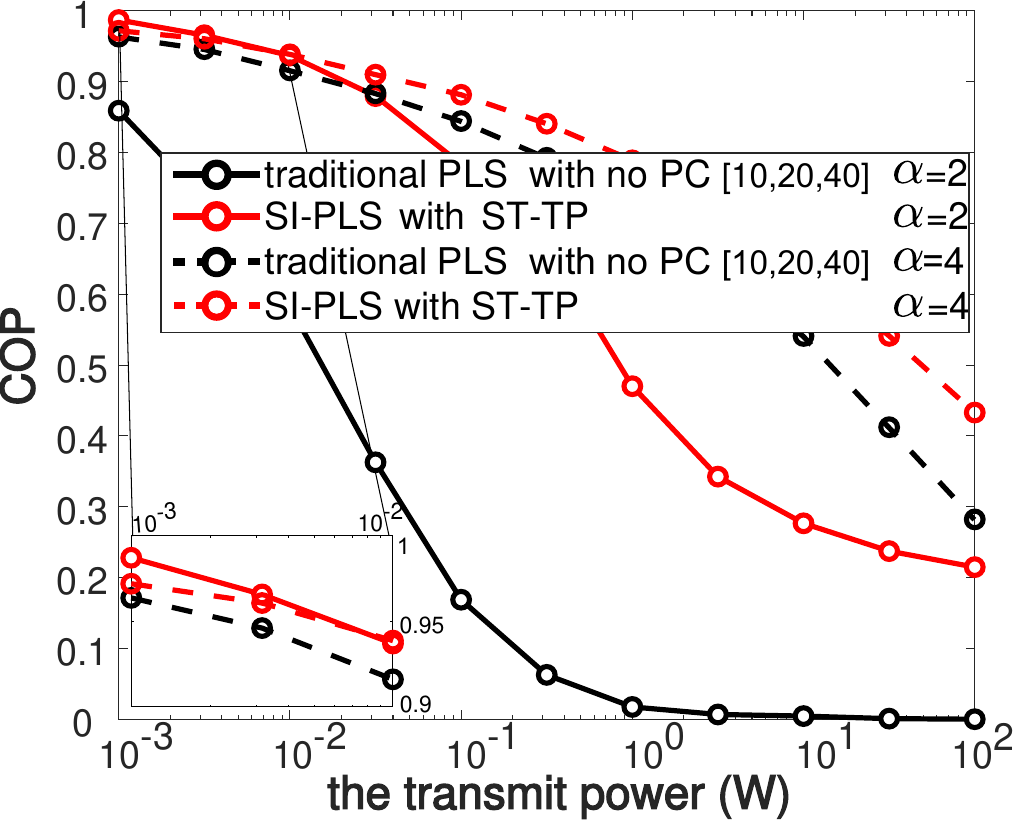}
  \caption{\small COP vs. commun. power}
  \label{fig:COP-P}
\end{minipage}
  \hfill
\begin{minipage}{0.24\textwidth}
        \centering
  \includegraphics[width=1.55in]{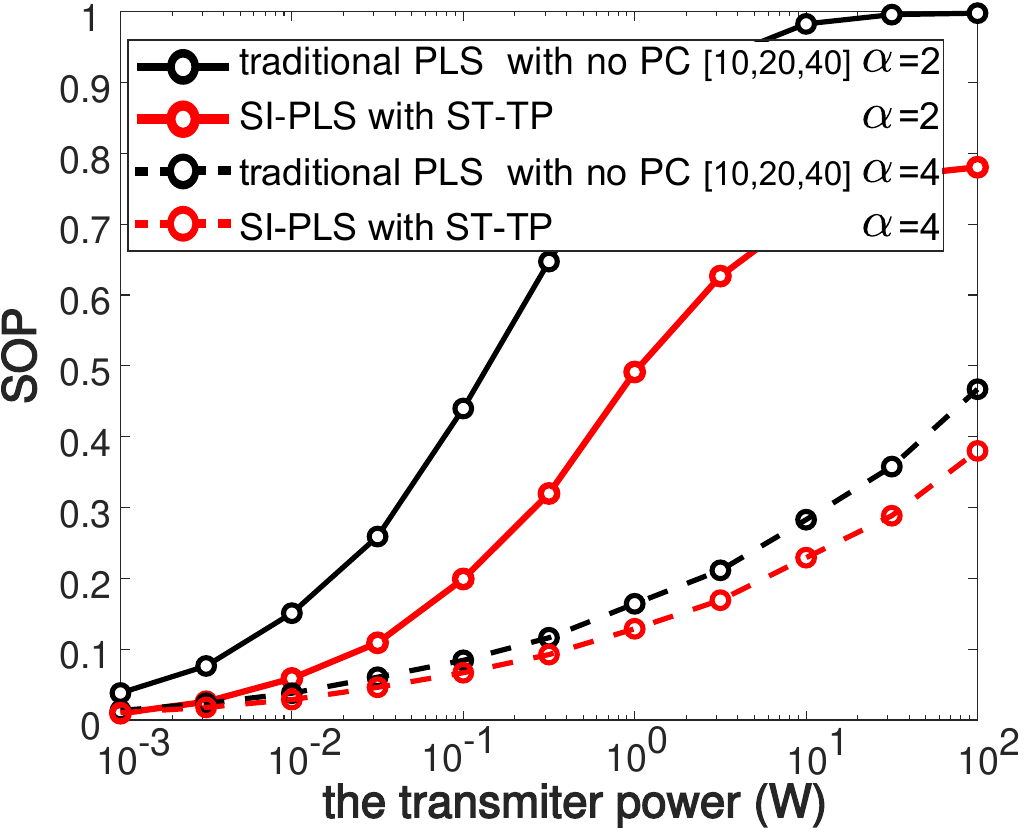}
  \caption{\small SOP vs. commun. power}
  \label{fig:SOP-P}
\end{minipage}
\end{figure*}

\begin{figure*}[htbp]
    \centering
\begin{minipage}{0.24\textwidth}
        \centering
\includegraphics[height=1.3in,width=1.5in]{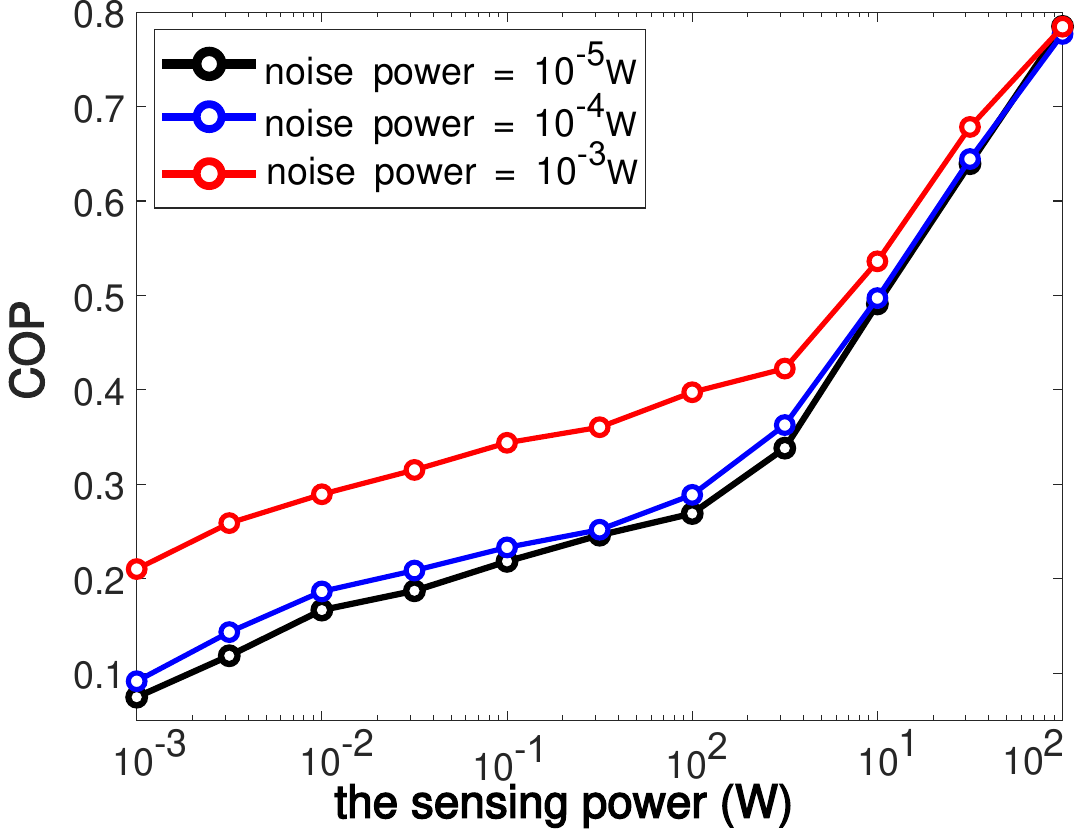}
  \caption{\small COP vs. sensing power}
  \label{fig:COP-P0}
\end{minipage}
  \hfill
\begin{minipage}{0.24\textwidth}
  \centering
  \includegraphics[width=1.55in]{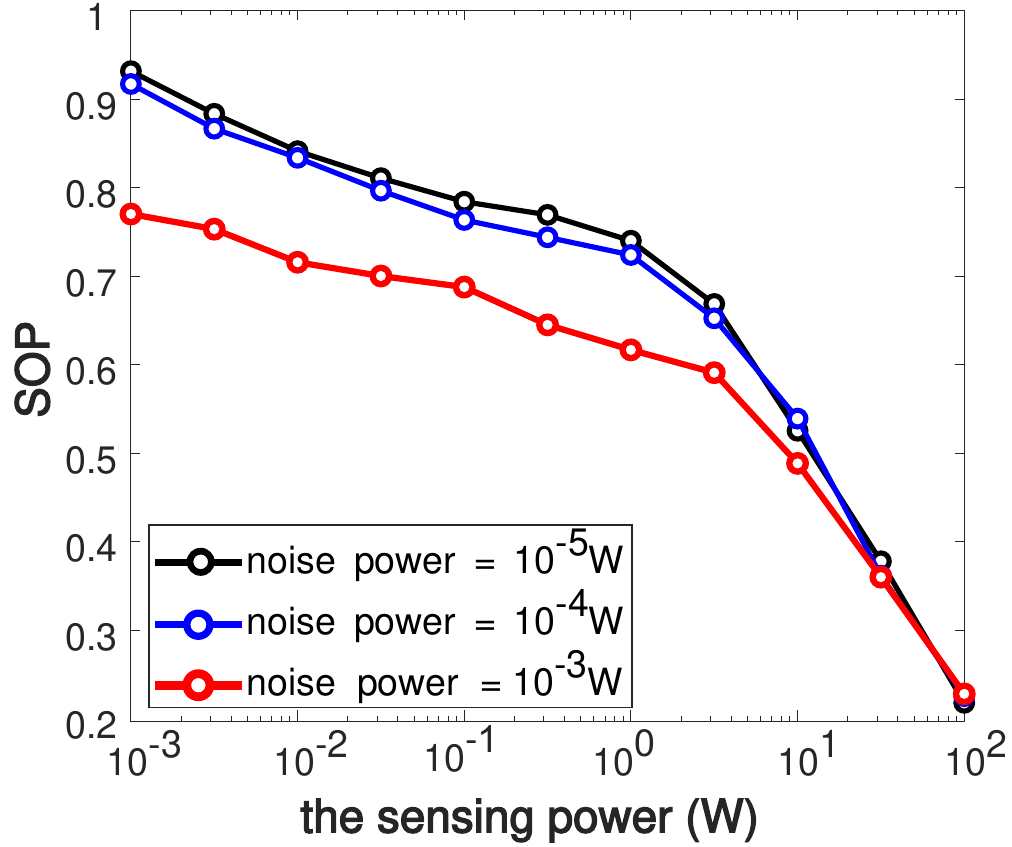}
  \caption{\small SOP vs. sensing power}
  \label{fig:SOP-P0}
\end{minipage}
  \hfill
\begin{minipage}{0.24\textwidth}
  \centering
  \includegraphics[width=1.55in]{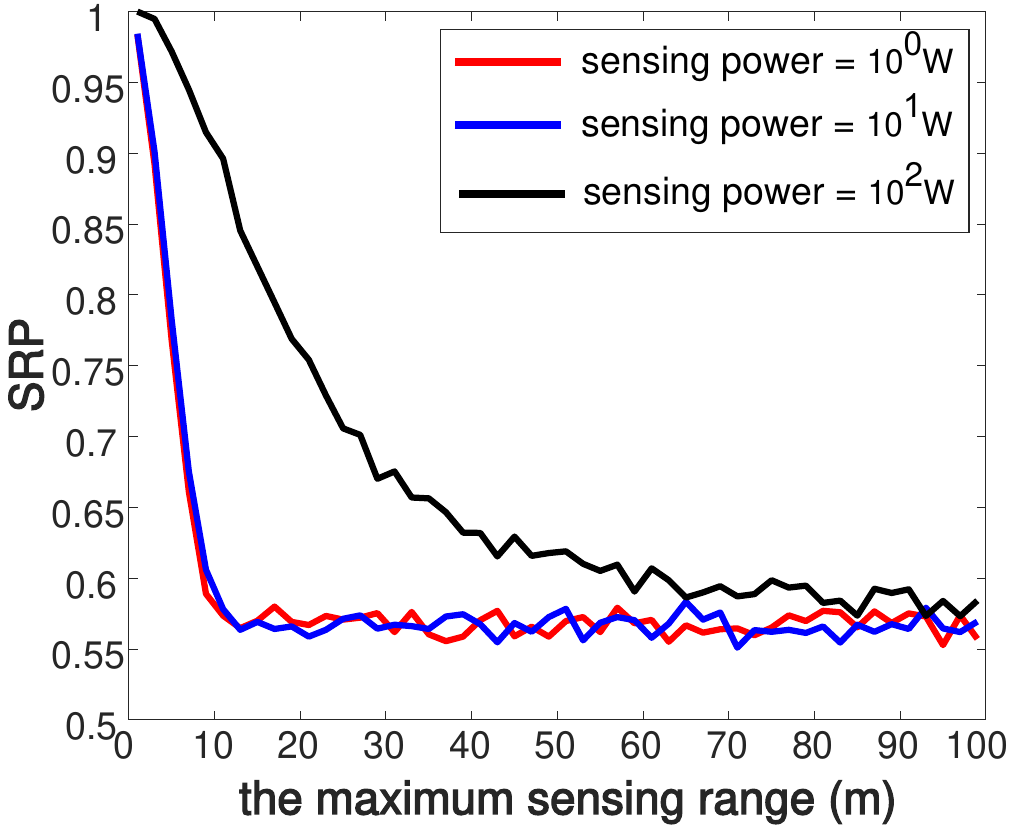}
  \caption{\small SRP vs. sensing range}
  \label{fig:SRP_Range(detectionpower)}
\end{minipage}
  \hfill
\begin{minipage}{0.24\textwidth}
	\centering
	\includegraphics[width=1.55in]{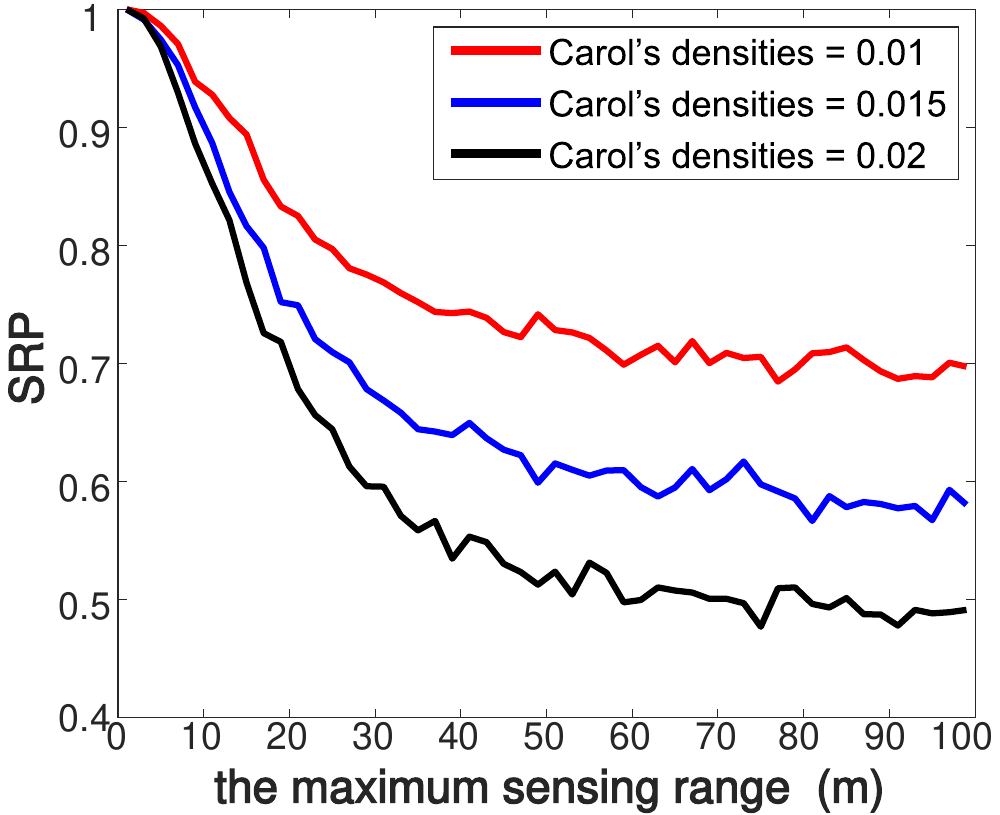}
	\caption{\small SRP vs. sensing range}
	\label{fig:SRP-Density}
\end{minipage}
\end{figure*}

First, we consider the performance comparison of \textbf{SI-PLS with ST-TP}, \textbf{traditional PLS with no PC} \cite{yin2021uav,jin2024enhanced,Li2024Joint} and with multiple Carols and Eves. We analyze the effects of the noise power, communication/sensing power, and radar horizontal beamwidth on communication reliability and security, as well as the sensing accuracy, as shown in Fig. \ref{fig:COP-Noise} to Fig. \ref{fig:Rs-P0}.

Fig. \ref{fig:COP-Noise} and Fig. \ref{fig:SOP-Noise} illustrate the variations in COP and SOP with the noise power under different settings of the path loss conditions. The results indicate that as the noise power increases, COP gradually increases, while SOP decreases. This is because that the noise deteriorates the reliability performance in the sense that the Bob cannot recover messages successfully, and simultaneously helps the security performance in the
sense that Eves cannot recover messages successfully.
In addition, higher path loss results in the weaker signal strength received by Bob and Eves, decreasing reliability and improving security. The corresponding reasons are similar to these of the noise power. Compared with the \textbf{traditional PLS with no PC} \cite{yin2021uav,jin2024enhanced,Li2024Joint}, on the average, the COP and SOP of \textbf{SI-PLS with ST-TP} can be decreased by $82\%$ and $41\%$ for the settings of $\alpha=2$, respectively. 
This phenomenon indicates that the proposed scheme performs better in terms of secure communications, particularly under conditions of a lower noise power and higher path loss. 

The impact of the transmit power on transmission reliability and security is shown in Fig. \ref{fig:COP-P} and Fig. \ref{fig:SOP-P}, respectively. The results indicate that increasing transmit power can achieve better reliability but gives a poorer security, which further corroborates the findings in Fig. \ref{fig:COP-Noise} and Fig. \ref{fig:SOP-Noise}. That is, for \textbf{traditional PLS with no PC} \cite{yin2021uav,jin2024enhanced,Li2024Joint}, a higher path loss exponent contributes to an improved security but a compromised reliability, as this scheme ignores the radar sensing interference. While for \textbf{SI-PLS with ST-TP}, when the transmit power is lower than $10^{-2}\rm{W}$, a greater path loss exponent improves reliability by reducing radar sensing interference from the Carols to the Bob. However, the communication signal is more susceptible to path loss exponent at higher levels of transmit power, necessitating the mitigation of smaller path losses. For the settings of a high path loss, on the average, the COP and SOP of the \textbf{SI-PLS with ST-TP} are reduced by $50\%$ and $35\%$, respectively, when compared to that of the \textbf{traditional PLS with no PC} \cite{yin2021uav,jin2024enhanced,Li2024Joint}. It is evident that traditional paradigms ignore the sensing interference signals, which consequently results in a decline in reliability when such interference is taken into consideration. However, this deficit can be effectively mitigated through the optimization of vehicle trajectory and communication power.

To further validate the impact of the sensing power, Carols density, and radar horizontal beamwidth on the performance of transmission reliability and security achieved by \textbf{SI-PLS with ST-TP}, Fig. \ref{fig:COP-P0} and Fig. \ref{fig:SOP-P0} consider the impact of the radar sensing power transmission reliability and security, respectively. It can be observed that increasing sensing power benefits the security enhancement, as the Carols provide more interference to the Eves, but it compromises reliability performance. The reasons are similar to these in Fig. \ref{fig:COP-Noise} and Fig. \ref{fig:SOP-Noise}. 

Finally, the influence of sensing range with different settings of sensing powers and Carols' densities on the sensing accuracy (i.e., SRP) are shown in Fig. \ref{fig:SRP_Range(detectionpower)} and Fig. \ref{fig:SRP-Density}, respectively. The results reveal that as the sensing range increases, the SRP experiences a significant decline. This is primarily caused by the severe degradation of the sensing signal over the distance increasing. For the cases of sensing power with 30dBm and 40dBm, the performance of sensing accuracy drops sharply when the detection range is below 10m. The reason behind this phenomenon is that the decline becomes slower once the detection distance increases, since the increased interference between the sensing signals that affects the overall sensing performance. In addition, there will be an increased level of interference to the sensing signal due to the presence of a greater number of interferers.

\subsection{Optimization of $TRSA\_SR$ maximization}
\begin{figure}[h]
\centering
\includegraphics[width=3in]{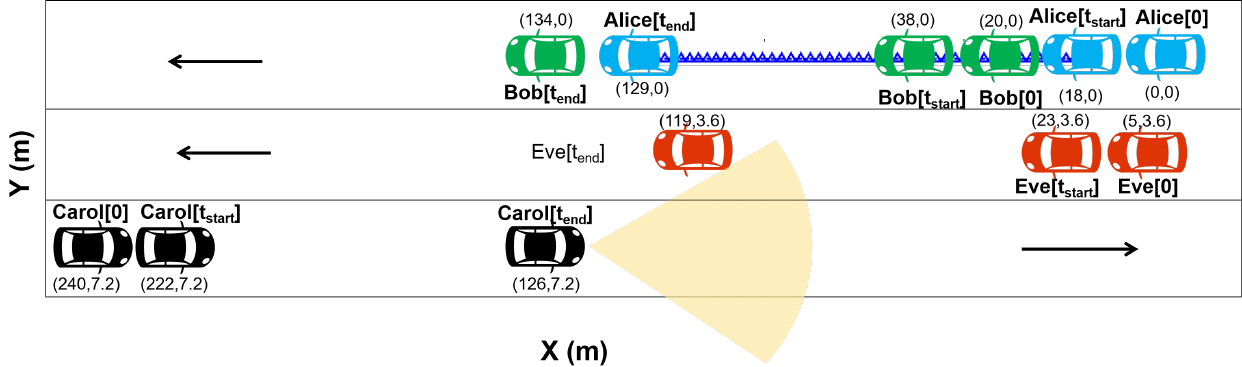}
\caption{\small vehicle positions in different time slots}
\label{fig:Position}
\end{figure}

To better observe the performance of \textbf{SI-PLS with ST-TP} and simplify representations of the earliest transmit time and latest transmit time of the Alice, we consider a scenario with an Eve and a Carol, and all vehicles move in a straight line along the current road. The maximum speed of the Alice is 20 m/s, while other vehicles move at a constant speed of  16 m/s. As shown in Fig. \ref{fig:Position}, the initial position of the vehicles is at $t[0]$, the secrecy rate by jointly optimizing the trajectory and transmit power of the Alice between  $t[k_{\rm start}]$ and $t[k_{\rm end}]$.

\begin{figure}[h]
\centering
\includegraphics[width=2.2in]{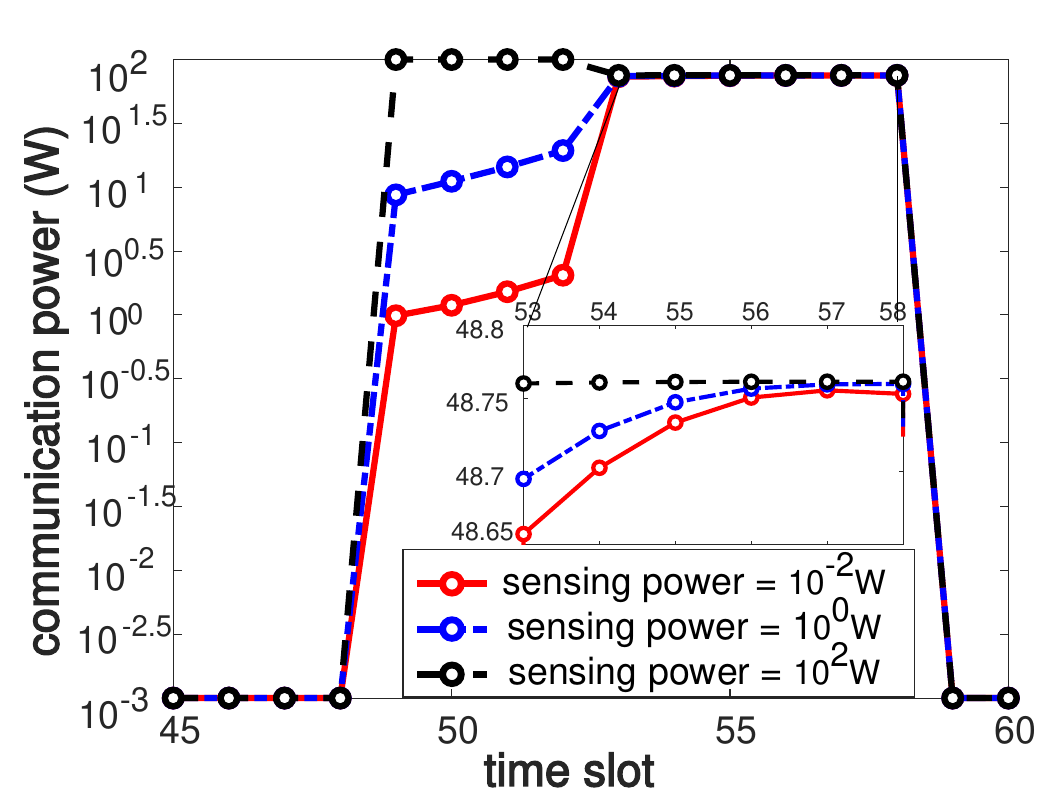}
\caption{\small communication power vs. different time slots}
\label{fig:OPT-P_T}
\end{figure}

\begin{figure}[h]
\centering
\includegraphics[width=2.2in]{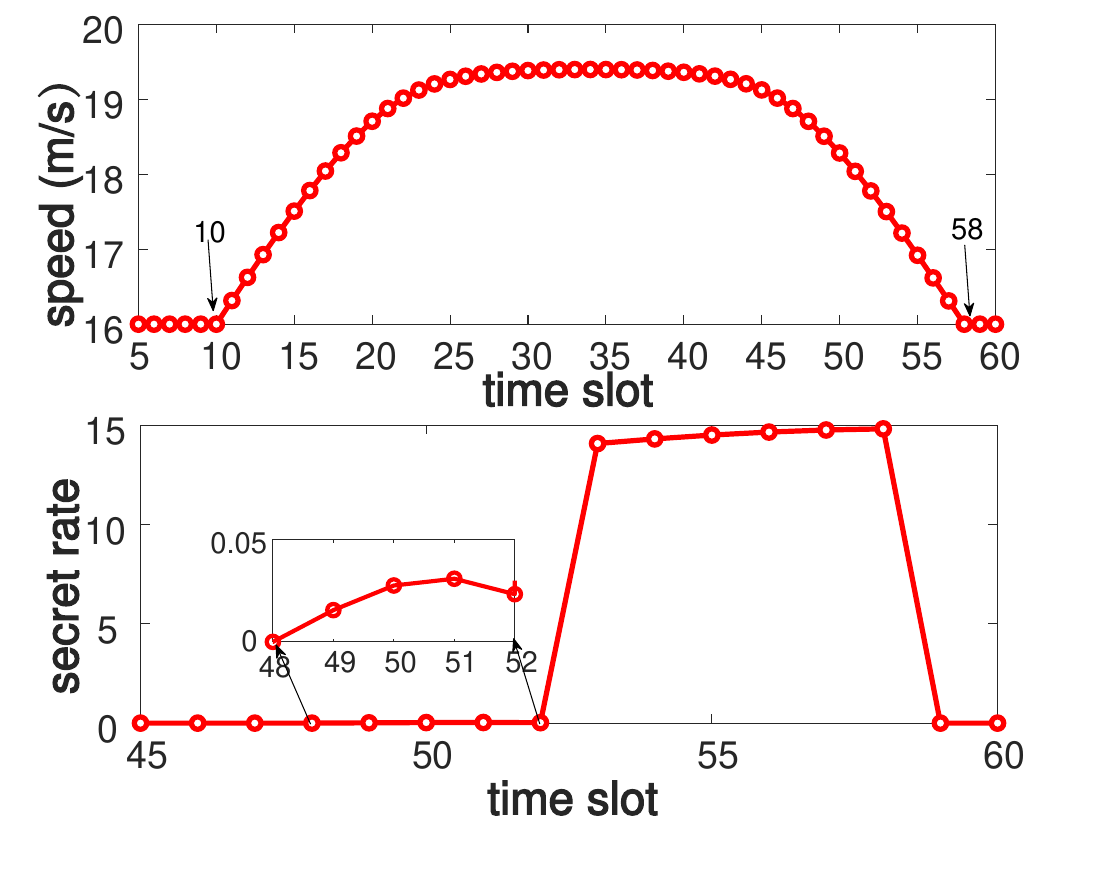}
\caption{\small moving speed and secrecy rate vs. different time slots}
\label{fig:OPT-V-Rs}
\end{figure}

At the beginning of 10-th time slot, the Eve suffers from the sensing interference generated by the Carol until 58-th time slot. During this time, the Alice optimizes the straight trajectory by adjusting its speed while keeping a safe distance with the Bob. Fig. \ref{fig:Position} and Fig. \ref{fig:OPT-P_T} describe the straight trajectory and transmit power of the Alice across different time slots. The variation of corresponding secrecy rate is shown at the bottom of Fig. \ref{fig:OPT-V-Rs}. It can be noticed from the top of Fig. \ref{fig:OPT-V-Rs} that  
although the Alice begins to optimize its trajectory and transmit power at the $10$-th time slot, the wiretap channel quality remains superior to that of the Bob, resulting in poor signal security, and then the transmit power is set to 0. From the $49$-th time slot to the 52-th time slot, by optimizing the straight trajectory and transmit power, the Bob obtains a better channel quality than that of the Eve. But the achieved secrecy rate is relatively lower, since the sensing interference affects the Bob and Eve simultaneously. Furthermore, the achieved secrecy rate rises rapidly at 53-th time slot, since the Bob moves out the radar sensing range of the Carol, only the Eve suffers from its sensing interference, and the straight trajectory and transmit power are also optimized. At the 58-th time slot, the Eve moves out the radar sensing range of the Carol, under which the Alice stops transmitting the confidential information and then the secrecy rate gets to zero.

\begin{figure}[h]
\centering
\includegraphics[width=2.2in]{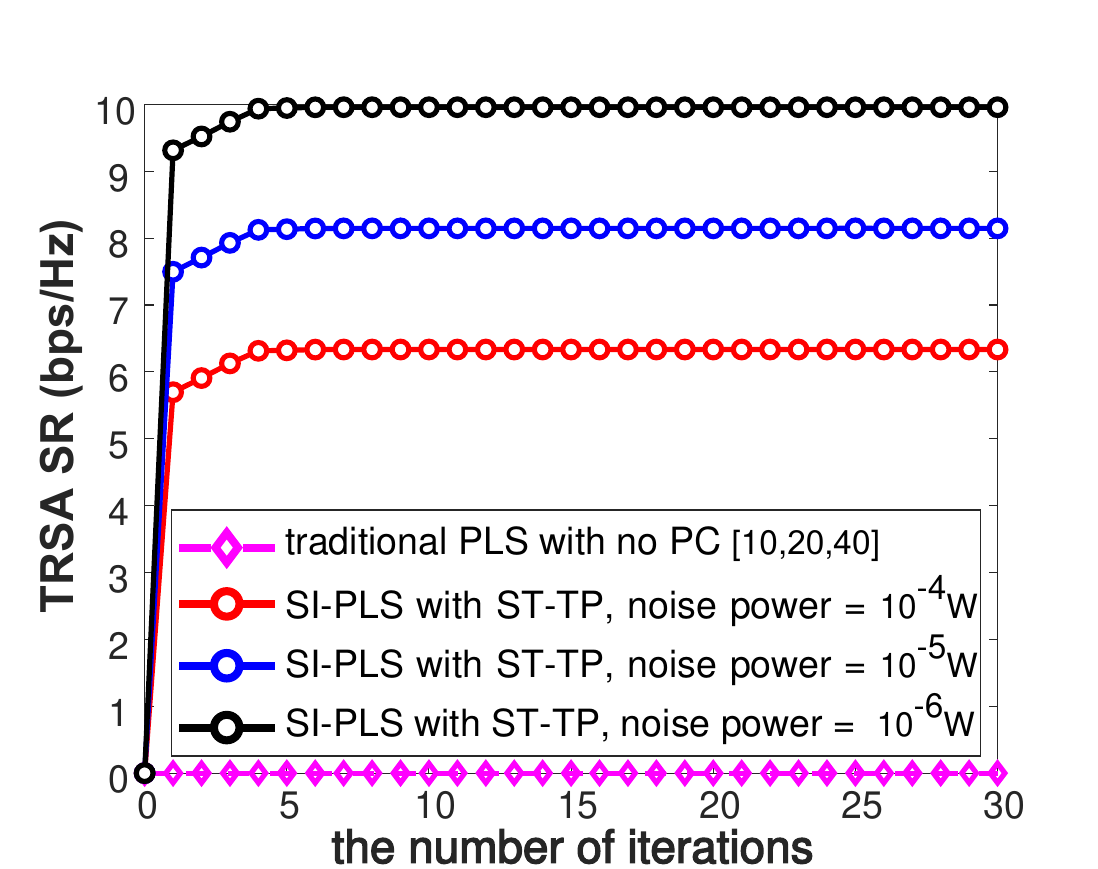}
\caption{\small $\rm TRSA\_SR$ vs. iterations with different noise powers}
\label{fig:OPT-iter}
\end{figure}

Fig. \ref{fig:OPT-iter} shows the convergence curve of the $\rm TRSA\_SR$ over the number of iterations. The secrecy rate achieved by using \textbf{traditional PLS with no PC} \cite{yin2021uav,jin2024enhanced,Li2024Joint} is zero because the Eve is closer to the Alice and has a better channel quality than that of the Bob. For \textbf{SI-PLS with ST-TP}, it can be seen that before the initial stage when the number of iterations is less than 4, the average secrecy rate gradually increases. After executing 5 iterations, $\rm TRSA\_SR$ reaches a relative stable stage, which means that the algorithm has converged to an optimal solution. To sum up, it can be concluded that the proposed algorithm effectively solves problem \eqref{eq:opt0}.

\begin{figure}[h]
\centering
\includegraphics[width=2.2in]{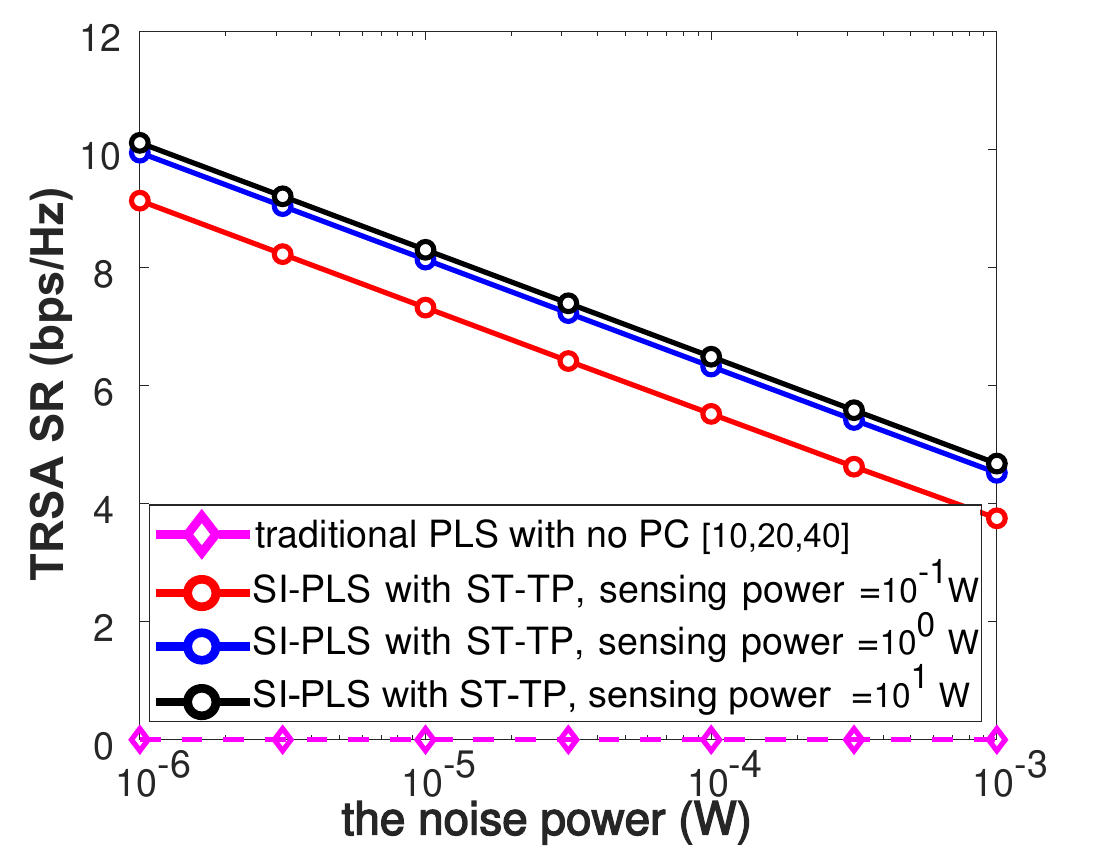}
\caption{\small $\rm TRSA\_SR$ vs. noise power with different sensing powers}
\label{fig:Rs-Noise}
\end{figure}

\begin{figure}[h]
\centering
\includegraphics[width=2.2in]{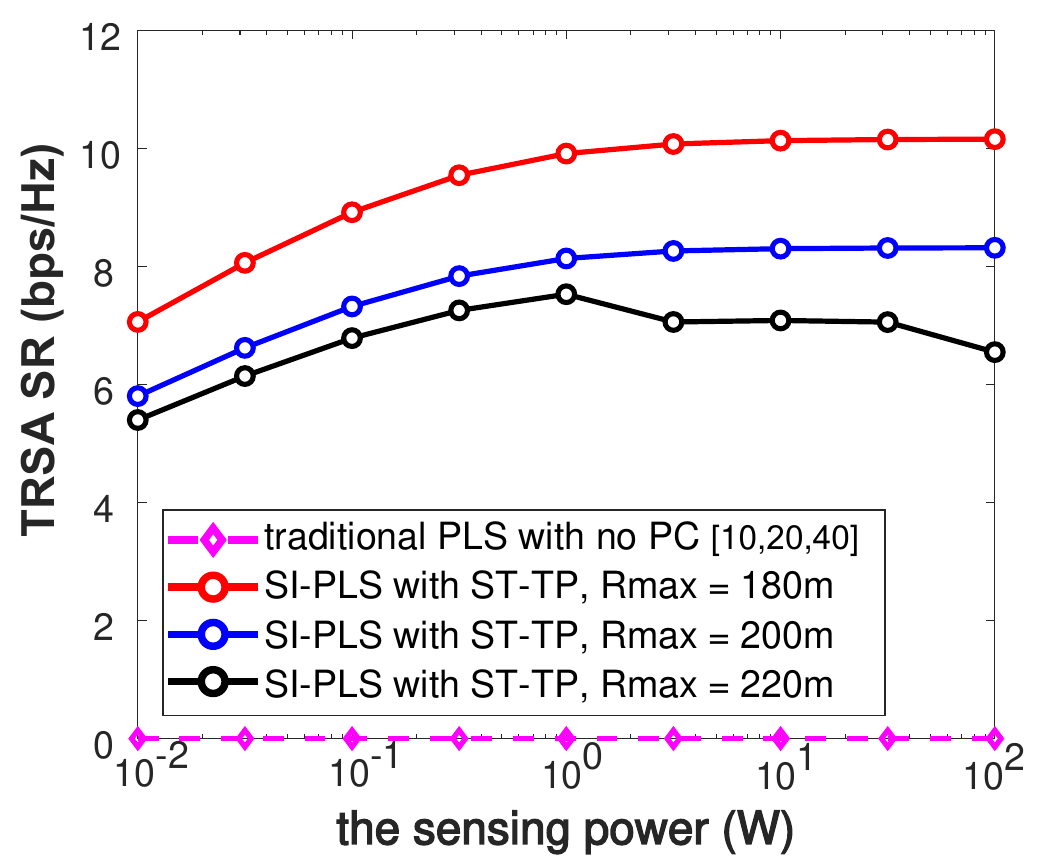}
\caption{\small $\rm TRSA\_SR$ vs. sensing power with different sensing ranges}
\label{fig:Rs-P0}
\end{figure}
Fig. \ref{fig:Rs-Noise} and Fig. \ref{fig:Rs-P0} present how average $\rm TRSA\_SR$ varies with different values of the noise power, radar sensing power and radar sensing range, respectively. It can be observed from Fig.\ref{fig:Rs-Noise} that, on the one hand, $\rm TRSA\_SR$ decreases over noise power increasing, since it affects the channel quality of the Bob more severely when compared with that of the Eve. On the other hand, a greater radar sensing power can obtain a higher $\rm TRSA\_SR$. The reason is that, via joint design of straight trajectory and transmit power, the Carol can suppress the Eve more heavily compared with the Bob, which also validates the effectiveness of our proposed \textbf{SI-PLS with ST-TP}. Furthermore, Fig. \ref{fig:Rs-P0} demonstrates this conclusion. In addition, we can observe that the radar sensing power should be selected properly for a greater sensing range. The reason is that although a greater radar sensing power can achieve a higher $\rm TRSA\_SR$, a greater radar sensing range with a greater sensing power can determinate the channel quality of the Bob more seriously, then $\rm TRSA\_SR$ decreases. Combining with the results in Fig. \ref{fig:Position}, the sensing accuracy and transmission security can be ensured by selecting a proper setting of the radar sensing power and radar sensing range.

\section{Conclusion}\label{sec:conclusion}
In this paper, we leverage the radar sensing interference to enable PLS of the IoV, rather than the communication interference, which is the core idea of traditional PLS techniques. The metrics for sensing accuracy, transmission reliability and security have been thoroughly evaluated. By jointly designing the transmit power and straight trajectory of the Alice, we formulate an optimization problem to maximize the secrecy rate of sensing interference enabled PLS. The optimization problem is non-convex and difficult to be solved directly, and then is decomposed two sub-problems and effectively addressed by using BCD and SCA methods. Furthermore, an AO algorithm is introduced to solve the original optimization problem. The feasibility of the proposed algorithm is verified by time complexity and convergence analysis. Simulations demonstrated the effectiveness of the proposed PLS method, and the impact of key system parameters on the sensing accuracy, transmission reliability and security.

\bibliographystyle{IEEEtran}
\bibliography{ref}
\end{document}